\documentclass[aps,pra,twocolumn,superscriptaddress,amsmath,amssymb,showpacs]{revtex4-1}

\usepackage{graphicx}	
\usepackage{dcolumn}	
\usepackage{bm}		
\usepackage{verbatim} 	
\usepackage{braket}
\usepackage{xr}
\usepackage{hyperref}

\begin{document}

\title{Percolation thresholds for photonic quantum computing}


\author{Mihir Pant}
\email{mpant@mit.edu}
\affiliation{Department of Electrical Engineering and Computer Science, MIT, Cambridge, MA 02139, USA}
\affiliation{Quantum Information Processing group, Raytheon BBN Technologies, 10 Moulton Street, Cambridge, MA 02138, USA}
\author{Don Towsley}
\affiliation{College of Information and Computer Sciences, University of Massachusetts, Amherst, MA 01003, USA}
\author{Dirk Englund}
\affiliation{Department of Electrical Engineering and Computer Science, MIT, Cambridge, MA 02139, USA}
\author{Saikat Guha}
\affiliation{Quantum Information Processing group, Raytheon BBN Technologies, 10 Moulton Street, Cambridge, MA 02138, USA}

\begin{abstract}
Any quantum algorithm can be implemented by an adaptive sequence of single node measurements on an entangled cluster of qubits in a square lattice topology. Photons are a promising candidate for encoding qubits but assembling a photonic entangled cluster with linear optical elements relies on probabilistic operations. Given a supply of $n$-photon-entangled microclusters, using a linear optical circuit and photon detectors, one can assemble a random entangled state of photons that can be subsequently ``renormalized" into a logical cluster for universal quantum computing. In this paper, we prove that there is a fundamental tradeoff between $n$ and the minimum success probability $\lambda_c^{(n)}$ that each two-photon linear-optical fusion operation must have, in order to guarantee that the resulting state can be renormalized: $\lambda_c^{(n)} \ge 1/(n-1)$. We present a new way of formulating this problem where $\lambda_c^{(n)}$ is the bond percolation threshold of a logical graph and provide explicit constructions to produce a percolated cluster using $n=3$ photon microclusters (GHZ states) as the initial resource. We settle a heretofore open question by showing that a renormalizable cluster can be created with $3$-photon microclusters over a 2D graph without feedforward, which makes the scheme extremely attractive for an integrated-photonic realization. We also provide lattice constructions, which show that $0.5 \le \lambda_c^{(3)} \le 0.5898$, improving on a recent result of $\lambda_c^{(3)} \le 0.625$. Finally, we discuss how losses affect the bounds on the threshold, using loss models inspired by a recently-proposed method to produce photonic microclusters using quantum dot emitters.
\end{abstract}
\maketitle

\section{Introduction}\label{sec:intro}

In {\em linear-optical quantum computing} (LOQC), a single photon in one of two orthogonal (spatial, temporal, or polarization) modes, i.e., $|10\rangle \equiv \ket{0}_L$ and $|01\rangle \equiv \ket{1}_L$ encodes a qubit, and passive linear optical interferometers and single-photon detectors are used to implement gates and measurements. Since each qubit is encoded by one photon, we use {\em photon} and {\em qubit} synonymously. Gates and measurements in LOQC are inherently probabilistic even if all single-photon sources are ideal and all linear optical elements and detectors are lossless. Component losses further reduce success probabilities, which translates into daunting requirements on number of devices (e.g., sources and detectors) to encode problems of practically-relevant size. Since the original Knill-Laflamme-Milburn (KLM) proposal for LOQC~\cite{2001.Nature.Knill-Milburn.KLM}---which was largely deemed unscalable due to the aforesaid reason---several variants have been proposed that use separately-prepared ``ancilla" states and photon number resolving (PNR) detectors to boost the probabilities of nondeterministic operations. 

A particularly promising variant is an LOQC architecture in the cluster-state model of quantum computing (QC)~\cite{2001.PRL.Raussendorf-Briegel.ClusterStateComp, 2005.PRL.Browne-Rudolph.Clusterstates}, which was introduced by Kieling, Rudolph and Eisert \cite{2007.PRL.Kieling-Eisert.PercolationQC, 2009.Book.Kieling-Eisert.Percolation}. This scheme leverages percolation and renormalization, to (a) probabilistically fuse many tiny microclusters (i.e., clusters of few entangled photons) using linear optical circuits into a randomly-grown large cluster, and (b) reinterpret the random instance of a large entangled cluster as a logical cluster state in the 2D square grid topology, which is a sufficient resource for universal QC \cite{2001.PRL.Raussendorf-Briegel.ClusterStateComp}. Rudolph and colleagues subsequently showed constructions within the above framework, which they termed {\em ballistic} photonic QC, wherein they demonstrated that with $3$-photon microclusters as an initial resource, one can create a percolated cluster with one-way transmission through a linear optical circuit, i.e., with no feedback \cite{2015.PRL.Gimeno-Segovia-Rudolph.3GHZtoBallisticQC, 2015.PRA.Zaidi-Rudolph.BallisticLOQC}. 

\begin{figure}[h]
    \includegraphics[width=\columnwidth]{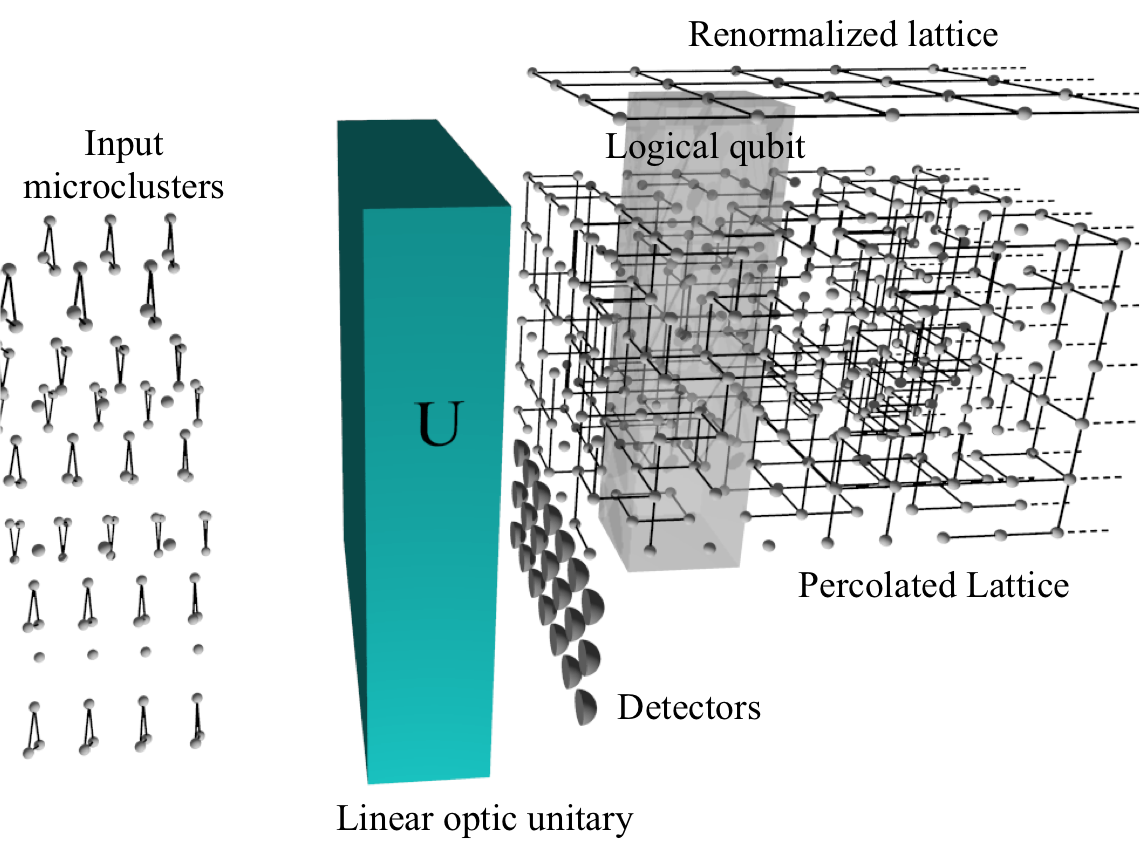}
    \caption{Ballistic photonic cluster state generation for quantum computing. A steady stream of entangled microclusters of size $n$-photons or less ($n=3$ shown) is incident on a linear-optical interferometer (i.e., a multimode unitary transformation $U$), which produces an entangled cluster of photons at its output. If a percolation condition is met, the output can be renormalized into a fully-connected logical cluster in a topology universal for cluster-model quantum computing.} 
    \label{fig:highlevelsetup}
\end{figure}
One can interpret the aforesaid feedback-free, or ballistic, framework of LOQC in the form visualized in Fig.~\ref{fig:highlevelsetup}, by `pushing out' (postponing) the detections involved in all the cluster-fusion operations to the very end. Consider an $N$-mode-input $N$-mode-output linear optical circuit---i.e., one that can be put together using $O(N^2)$ beamsplitters and phase-shifters \cite{1994.PRL.Reck-Zeilinger.DiscreteUnitaryopfrombeamsplitter}---and whose action on the input modes is described by a complex-valued unitary matrix $U$. At each time step, the linear-optical circuit is fed with several microclusters (of up to $n$ entangled photons each) that occupy $M$ of the $N$ input modes. As we show, if a certain percolation threshold is exceeded, the spatio-temporal entangled sheet of photons that emerges at the output of $U$ is a resource that is universal for cluster model quantum computing. This is true in the following sense. A fraction of the output modes is detected using PNR detectors at each time step. In the final ``renormalization" step, the entangled state that emerges in the remaining output modes is broken up into logical blocks using information from the PNR detection outcomes. Exactly one representative photon is left unmeasured in each logical block while the rest of the photons are measured in appropriate bases to leave the representative photons in each logical block in a fully-connected 2D square grid cluster, into which one can encode any quantum algorithm. We emphasize that the detection outcomes are only used for the renormalization step, i.e., to figure out how to use the randomly connected output cluster for QC; they are {\em not} used to determine whether or not the unmeasured part of the output cluster is useful for universal QC (this is true with near certainty if the percolation condition is met). 

A major open question which we address in this paper is---if $n$-photon microclusters are the input resource, what is the minimum probability of success $\lambda_c^{(n)}$ with which each two-photon fusion attempt must succeed, such that one is guaranteed a percolated renormalizable cluster for universal quantum computing, assuming that the best possible spatio-temporal sequence of two-photon fusion attempts are employed on the input microclusters. Entangled microclusters can be used to increase the fusion success probability beyond $0.5$, the highest value attainable with linear optics and photon detection alone~\cite{2011.PRL.Grice.BoostedBM, 2014.PRL.Ewert-Loock.boostedfusion}. Therefore, some of the input microclusters can serve as building blocks for the percolated cluster while others can be used to boost the fusions. Therefore, a second important open question is: what is the maximum success probability $\lambda_{\rm max}^{(n)}$ attainable with $n$-photon microclusters used as an ancillary resource? Clearly we need $\lambda_{\rm max}^{(n)} > \lambda_c^{(n)}$ for it to be possible to obtain a renormalizable percolated cluster. As $n$ increases, $\lambda_{\rm max}^{(n)}$ and $\lambda_c^{(n)}$ are likely to increase and decrease respectively, driving the percolated cluster deeper into the supercritical-connected regime, which makes the construction more efficient by driving the dimensions of the renormalized blocks (and hence the number of photons that map to one logical lattice node in the renormalized lattice) to be smaller. Furthermore, if one allows for simultaneous fusion of three or more photons, very little is known about success probabilities of linear optical fusion and it is not clear if the thresholds and the efficiency of the above construction improves.

Another important practical question is the effect of losses and other device imperfections on the ballistic creation of resources for universal QC. If $\eta \in (0,1)$ is the transmissivity each photon sees through its lifetime (including losses in the source, waveguides and detectors), as $\eta$ decreases from one (the ideal lossless limit), $\lambda_{\rm max}^{(n)}(\eta)$ decreases while $\lambda_c^{(n)}(\eta)$ increases. There is a threshold on $\eta_c^{(n)}$ such that if $\eta < \eta_c^{(n)}$, $\lambda_{\rm max}^{(n)} > \lambda_c^{(n)}$ is no longer true. An open question therefore is whether this loss tolerance threshold improves with increasing size of input microclusters (i.e., $\eta_c^{(n)}$ decreases as $n$ increases), and if so at what rate. Finally, in the presence of photon loss, since we don't know where losses occured, constructing a fully-connected universal logical renormalized cluster is non-trivial, and has not been addressed in the literature. To address this, one could modify the above scheme to start with the creation of logical photonic qubits that are tolerant to losses and other errors such as mode mismatch and detector noise, and thereafter do fusion, percolation and renormalization on these error-protected logical qubits. 

When restricted to $n=1$, i.e., only single photons as the initial resource, our setup in Fig.~\ref{fig:highlevelsetup} resembles that of {\em Boson Sampling} (BS), a physics-based computation model introduced and analyzed by Aaronson and Arkhipov \cite{2011.ProcACM.Aaronson-Arkhipov.linear_optics_complexity, 2013.TheoryofComputing.Aaronson-Arkhipov.linear_optics_complexity}. If $M$ photons are fed into a randomly chosen linear optical circuit $U$, and if {\em all} the output modes are detected using PNR detectors, the setup naturally samples from the induced $N$-mode $M$-photon joint probability mass function (pmf) at the output of $U$. It was shown that drawing samples from this particular joint pmf is very likely not possible efficiently on a classical computer. However, it is also believed that BS does not have the computational power of universal quantum computing. The computational hardness of sampling from the output joint photon number distribution when $n \ge 2$ input clusters are employed, has not been analyzed. We emphasize however that the problem we described above (i.e., what conditions must be satisfied for the entangled state at the output of $U$ to be a resource that is sufficient for universal QC) is distinct from the problem at the heart of BS (the computational hardness of sampling from the joint photon number distribution of the entangled state at the output of $U$). It will however be interesting to explore if there is a closer connection between the two problems, and whether a connectivity metric on the output entangled state can be mapped in a meaningful way to computational hardness of sampling from its joint photon number distribution. 

\section{Main results}

Let us assume destructive two-photon fusion operations that succeed with probability $\lambda$. In other words, each fusion operation is assumed to act on two photons at a time, and regardless of whether the fusion succeeds or fails, those two photons are destroyed. With the optimal choice of sequence/pattern/algorithm to fuse the $n$-photon clusters, there exists an optimal (minimal) threshold $\lambda_c^{(n)}$, such that if all fusions succeed with probability $\lambda > \lambda_c^{(n)}$, the end product is a percolated cluster renormalizable for universal QC. These thresholds $\lambda_c^{(n)}$ for $n = 1, 2, \ldots$, and ways to achieve them, in particular for small values of $n$, are important questions that need to be answered in order to understand the resource-optimal way to realize photonic QC. 


The results in this paper can be summarized as follows:

\begin{enumerate}
\item {\bf{Converse}}---We prove: $\lambda_c^{(n)} \ge 1/(n-1), \forall n \ge 2$, i.e., no matter how we choose to fuse $n$-photon clusters, if each two-photon fusion succeeds with probability less than $1/(n-1)$, the final cluster produced is fragmented with high probability, and not suitable for renormalization. This means that with $n=3$ microclusters (three-photon GHZ states) as the initial resource (as in \cite{2015.PRL.Gimeno-Segovia-Rudolph.3GHZtoBallisticQC, 2015.PRA.Zaidi-Rudolph.BallisticLOQC}), the minimum $\lambda$ needed for percolation is $0.5$. With $n=2$ microclusters (Bell states) as the initial resource, if the fusions succeed with any probability less than one, the output cluster is not percolated. Hence, with pairwise destructive fusions, $n=3$ microclusters are the minimum size needed for ballistic LOQC. However, our converse does not immediately tell us whether there exists a systematic prescription to achieve percolation {\em at} $\lambda_c^{(n)} = 1/(n-1)$. We also show that if $m \ge 2$ node fusion operations are employed to fuse $n$-qubit microclusters, the percolation threshold must satisfy: $\lambda_c^{(n, m)} \ge 1/\left[(n-1)(m-1)\right]$. However, very little is known about linear-optical circuits for $m > 2$ qubit fusion~\cite{1998.PRA.Pan-Zeilinger.GHZanalyzer} (e.g., projecting $3$ qubits to one of the $8$ three-qubit GHZ states) and their associated success probabilities. Therefore, it remains unclear if the above bound on $\lambda_c^{(n, m)}$ is tight.

\item {\bf{New percolation framework}}---We develop a new percolation framework to address the problem of assembling a large photonic cluster using cluster fragments, where the threshold on fusion success probability $\lambda_c^{(n)}$ maps on to the usual bond percolation threshold $p_c(G)$ of an appropriately defined logical graph $G$ each of whose nodes corresponds to an $n$-photon microcluster. Each node in $G$ is assigned a color based on how many of the $n$ photons in the microcluster at that node are intended to be measured in fusion attempts, which is the node's degree, whereas each fusion attempt corresponds to a neighboring bond of a node in the logical graph. 

\item {\bf{Improved achievable fusion thresholds}}---
Using our percolation framework, we present new constructions and associated fusion success thresholds for percolation. The lowest threshold we show achievable with $n=3$ microclusters is $\approx 0.5898$ which improves over a recently published threshold of $0.625$~\cite{2015.PRL.Gimeno-Segovia-Rudolph.3GHZtoBallisticQC}. 

\item {\bf{Ballistic percolated cluster generation with a 2D graph}}---We show a logical graph construction using a modified brickwork lattice, with which it is possible to fuse $3$-photon microclusters in a 2D (planar) topology and achieve percolation at $\lambda_c \approx 0.746$. This threshold being less than $0.78125$ makes it possible to achieve using single-photon boosted linear optical fusion~\cite{2014.PRL.Ewert-Loock.boostedfusion}. A planar architecture is very promising from an experimental standpoint because a planar integrated photonic waveguide can be used to weave such a cluster. This also shows it is possible to percolate a 2D lattice using single-photon-boosted fusion, a question left open by Rudolph~\cite{2016.arXiv.Rudolph.SiPhOptimism}. 

\item {\bf{Conjectured achievable thresholds with two-photon fusion}}---Finally, we conjecture, and provide compelling evidence in its favor, that if there is an infinite lattice $G$ of maximum node degree $n$ with bond percolation threshold $p_c$, it is possible to stitch together a giant percolated cluster renormalizable for QC using $n$-photon microclusters as long as the fusions succeed with probability $\lambda > p_c$. We show that the truth of this conjecture would imply that for $n=3$, the lowest known achievable threshold would go down to $0.5$, thereby proving $\lambda_c^{(3)} = 0.5$. We also conjecture, using an extension of the argument for $n=3$, that the converse bound we prove is tight, i.e., it is possible to construct a logical graph that can be percolated with two fusion success probability, $\lambda_c^{(n)} = 1/(n-1)$. 

\item {\bf{Loss tolerance of percolation thresholds}}---Using a photon loss model inspired by a recent proposal to produce photonic microclusters using quantum dot emitters~\cite{2009.PRL.Lindner-Rudolph.ClusMachGun, 2016.Science.Schwartz-Gershoni.QDClusGen}, we prove an extension of our converse result, i.e., we show a lower bound on $\lambda_c^{(n)}$ that is a function of $n$ and $\eta$ (a parameter that quantifies the loss experienced by each photon). In other words, if the two-photon fusion success probability is less than this lower bound, for no sequence of fusing photons with a collection of $n$-photon microclusters, can one get a renormalizable percolated cluster. We discuss the implications of our results to the loss tolerance of photonic quantum computing using this scheme. We also discuss important open problems that need addressing, primarily that of renormalizing a cluster in the presence of photon loss and other device imperfections.
 \end{enumerate}

\section{Revisiting ballistic cluster-state LOQC with a new approach}\label{sec:newpicture}

\begin{figure*}
    \includegraphics[width=0.8\textwidth]{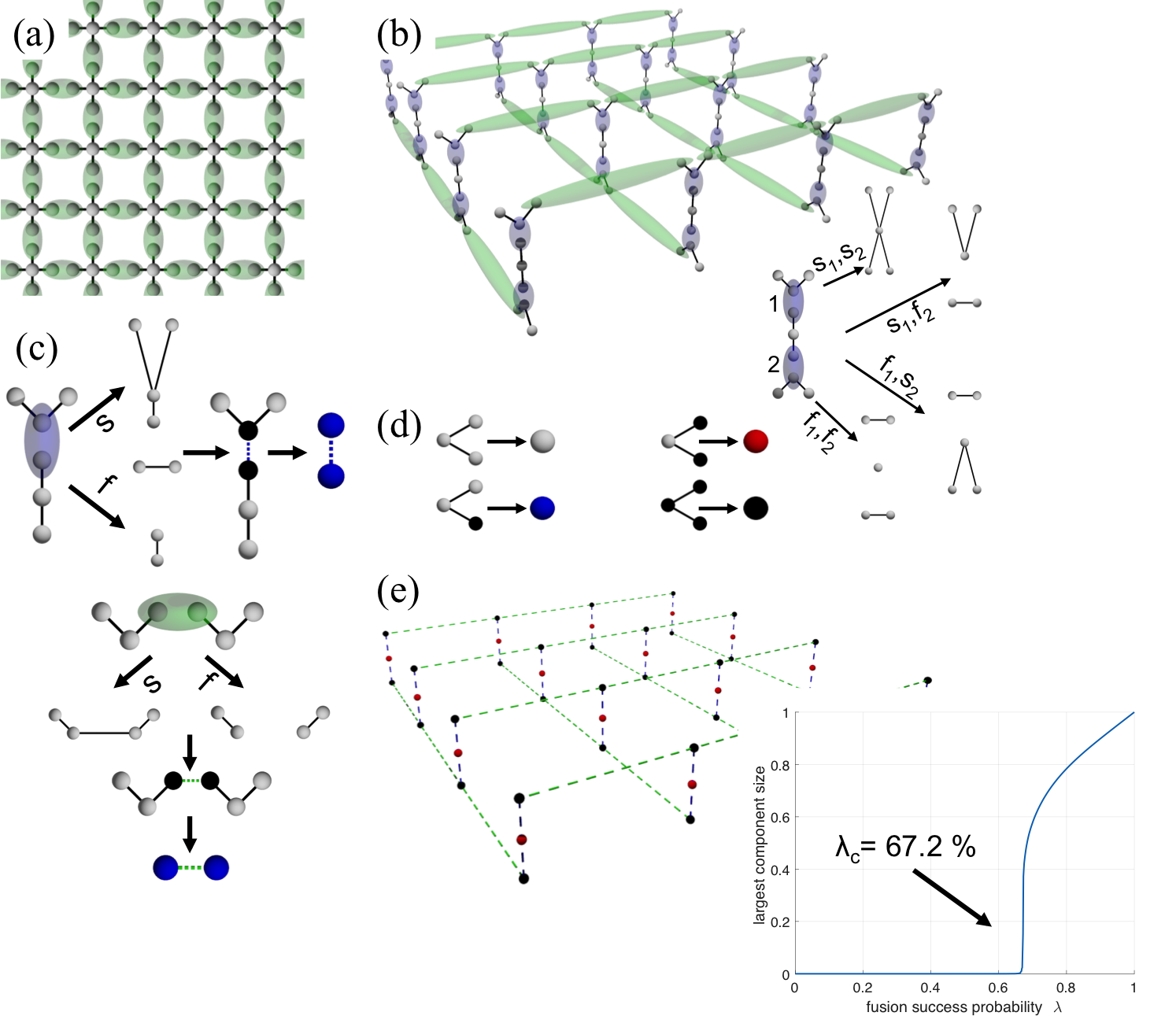}
    \caption{Different strategies and logical interpretations of piecing together a 2D square lattice by fusing microclusters: (a) $5$-photon microclusters at each lattice node with fusion attempts on each lattice bond; (b) vertical arrangements of three $3$-photon microclusters and $2$ fusions create a $5$-photon cluster if both fusions succeed; (c) interpreting fusion as coloring the measured nodes black and drawing a new bond between them if fusion succeeds, the linear optical circuits corresponding to the blue and green ellipses are shown in Fig. 2 and 3 of \cite{2015.PRL.Gimeno-Segovia-Rudolph.3GHZtoBallisticQC} respectively; (d) mapping microclusters to nodes in a logical graph and coloring them based on how many photons in the microcluster are left unmeasured; (e) pure bond percolation on the logical graph of colored nodes.} 
    \label{fig:Keilingtonow}
\end{figure*}

In this section, we develop a conceptually new way to construct percolated instances of renormalizable photonic clusters, and re-interpret recent results within our framework. We close the section with a conjecture. In Section~\ref{sec:fundthresholds} we use our percolation framework to develop new results on better achievable percolation thresholds, as well as general bounds on $\lambda_c^{(n)}$.

\subsection{Graph states and linear optical fusion}\label{sec:graphstateintro}

We consider clusters of entangled photons in this paper that belong to a special class called {\em graph states}~\cite{2001.PRL.Briegel-Raussendorf.ClustStateIntro}. A cluster described by the graph $G(V, E)$ can be prepared by placing one qubit in the state $(\ket{0}_L + \ket{1}_L)/\sqrt{2}$ at every node in $V$ and applying a two-qubit controlled phase (CZ) unitary operation across every edge in $E$. With single photons as the starting point, using passive linear optical circuits, a $2$-qubit cluster can be generated with a success probability of $3/16$ \cite{2008.PRA.Zhang-Pan.HeraldedBellPairsource}, and a $3$-qubit cluster (in line or triangle topology) can be generated with a success probability of $1/32$ \cite{2008.PRL.Varnava-Rudolph.PhotonLoss-LOQCscaling}, both assuming lossless linear optics and ideal detectors. The maximum success probability of linear-optical two-photon fusion, $\lambda$ is $0.5$ when no ancilla photons are used \cite{1996.PRA.Michler-Zeilinger.BellStateAnalysis, 2001.AppPhysB.Calsamiglia-Lutkenhaus.BSMlimit}. Ancilla single photons can be used to achieve $\lambda = 0.78125$~\cite{2014.PRL.Ewert-Loock.boostedfusion}.

\subsection{Fusing microclusters on a regular lattice}\label{sec:fusingclustersonlattice}

\begin{figure}
    \includegraphics[width=\columnwidth]{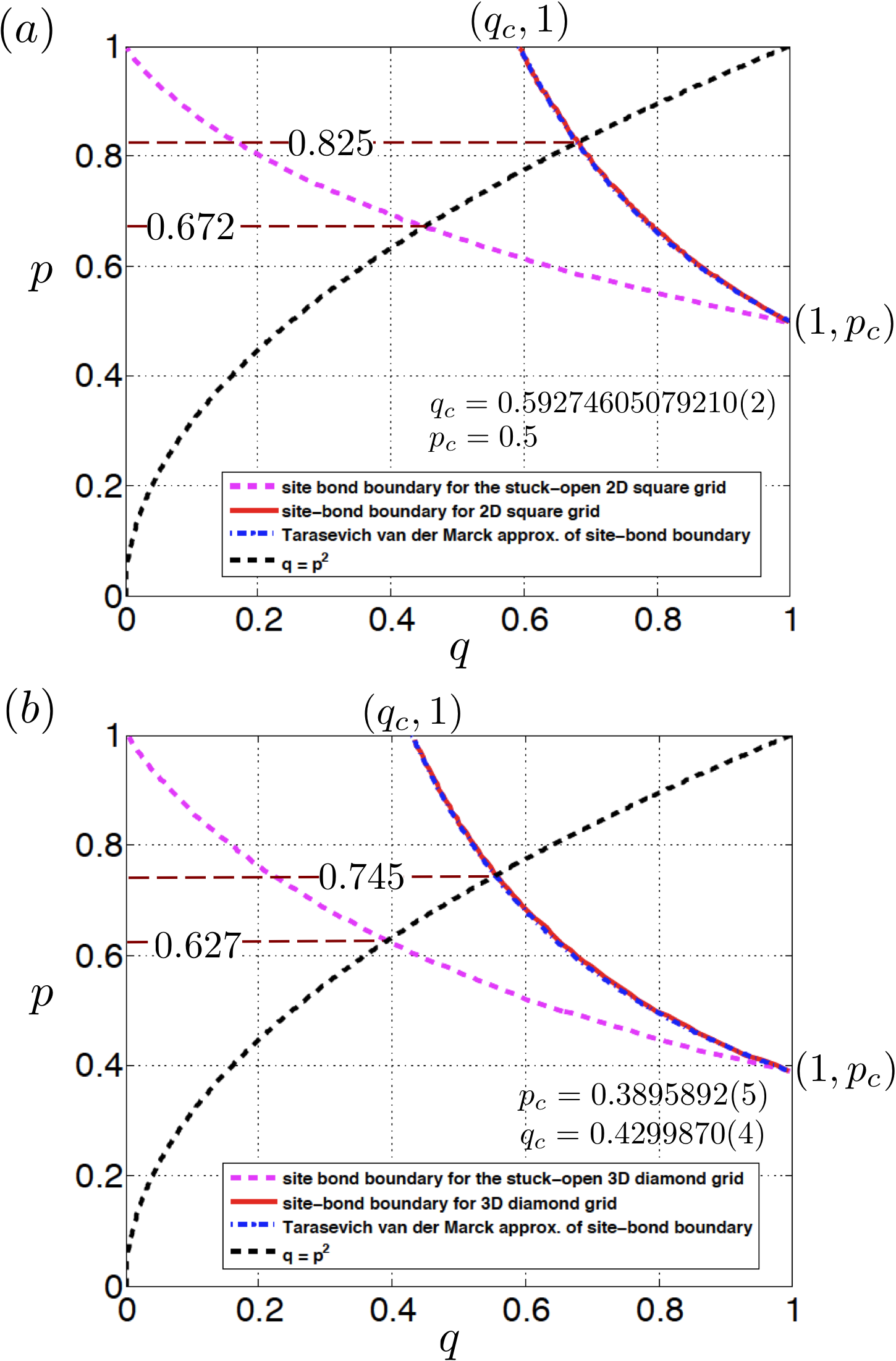}
    \caption{Site-bond percolation critical boundaries shown for the (a) 2D square and (b) 3D diamond lattices. The magenta curves correspond to a modified site-bond percolation problem described in the text where even if a site is not occupied, neighboring bonds can still be pairwise connected if occupied.} 
    \label{fig:sitebondregion}
\end{figure}
We begin with an illustrative example of piecing together a large subgraph of the 2D square lattice by probabilistic fusion of microclusters using two-photon destructive fusion operations that succeed with probability $\lambda$. 

\subsubsection{A conservative approach: site-bond percolation}\label{sec:Kielingsquare}

We begin by preparing $5$-photon clusters in a star topology and placing them at each node of the lattice, as shown in Fig.~\ref{fig:Keilingtonow}(a) \cite{2007.PRL.Kieling-Eisert.PercolationQC}. Suppose we succeed in preparing each of those clusters with probability $q$. We then attempt $2$-photon fusions across each edge of the lattice, each of which succeeds with probability $\lambda$. The resulting graph state that is generated is a random instance of site-bond (mixed) percolation \cite{1980.MathProcCambPhysSoc.Hammersley.sitebondperc} where each bond is occupied with probability $p = \lambda$ and each site is occupied with probability $q$. The boundary in the $(q, p)$ space that separates the percolating from the non percolating region is shown by the red solid plot in Fig.~\ref{fig:sitebondregion}(a). We also show an analytical approximation of this critical boundary (blue dash-dotted plot), developed by Tarasevich and van der Marck \cite{1999.IJMPC.Tarasevich-vandermarck.sitebondapprox}. If one had $3$-photon clusters (GHZ states) as a starting resource, one can assemble a $5$-photon star by attempting two fusions on three $3$-photon clusters, as shown in Fig.~\ref{fig:Keilingtonow}(c). The probability of success in creating the $5$-node star is thus $q = \lambda^2$, the probability that both fusions succeed. If either fusion fails, we call it a node failure. Therefore, per Kieling {\em et al.}'s recipe, the threshold value of $\lambda$ beyond which one gets percolation is given by the intersection of the site-bond critical boundary and the line $q=p^2$, thereby obtaining $\lambda_c \approx 0.825$ (see Fig.~\ref{fig:sitebondregion}(a)). 

\subsubsection{Exploiting failure modes: modified site-bond percolation with two stuck-open layers}\label{sec:Mercedessquare}

It is too conservative to ask for both fusions to succeed at every node \cite{2015.PRL.Gimeno-Segovia-Rudolph.3GHZtoBallisticQC}. In other words, even if one or both fusions in creating the $5$-node star were to fail, the leftover cluster fragments can still provide some connectivity. We illustrate this in Fig.~\ref{fig:Keilingtonow}(b), where we lay out the three $3$-photon clusters at each node of the square lattice in the vertical arrangement shown, while the square lattice is divided into two crisscrossing layers of parallel 1D lattices. It is as if the lattice is stuck open at each node. If both fusions at a node---shown as light blue ovals---succeed (which happens with probability $q = \lambda^2$), the photon at the center of the vertical arrangement gets attached to the two photons in the top layer as well as the two in the bottom layer, thus forming the $5$-photon star. This has the effect of connecting the two layers at that node. If one or both fusions at a node fail (with probability $1-q$), the node remains stuck open. But, even so, the two nodes in the top layer of the vertical arrangement remain connected to one another, and the same is true for the two nodes in the bottom layer. If one of the two fusions in the vertical arrangement succeeds (and the other fails), the two nodes in the layer closer to the successful fusion are connected via the center node, whereas the two nodes in the other layer (one closer to the failed fusion) are connected to one another directly. In all of these cases (i.e., if one or both fusions fail), the middle node plays no role in terms of providing long-range connectivity. The green ovals show fusion attempts between adjacent nodes in each of the two layers, the bonds of the square lattice.

The situation looks identical to $(q, p)$ site-bond percolation with $q = \lambda^2$ and $p = \lambda$, except that even if a site is not active, the four neighboring bonds at that site can be pairwise connected to one another in the two stuck-open layers. We numerically evaluated the percolation region of this modified site-bond problem using the Newman-Ziff algorithm \cite{2001.PRE.Newman-Ziff.NZAlgo} on a square lattice of $1$ million nodes. The resulting percolation boundary is shown in the magenta dashed plot in Fig.~\ref{fig:sitebondregion}(a). This intersects with $q=p^2$ at $p = \lambda \approx 0.672$. This threshold is already below $0.78125$, the success probability attainable by linear-optical fusion boosted with ancilla single photons \cite{2014.PRL.Ewert-Loock.boostedfusion}.

\subsubsection{Pure bond percolation on a logical graph}\label{sec:purebondsquare}

Let us revisit the picture in Fig.~\ref{fig:Keilingtonow}(b), and consider a new interpretation where each $3$-photon cluster is thought of as a single (super) node in a logical graph shown in Fig.~\ref{fig:Keilingtonow}(e). We assign a color to the super node based on how many of its photons are intended to be measured (and hence destroyed) in the planned fusion attempts (Fig.~\ref{fig:Keilingtonow}(d)). The central photons in the $3$-photon clusters at the centers of the vertical arrangements in Fig.~\ref{fig:Keilingtonow}(b) are not measured as part of a fusion. Hence, those $3$-photon clusters map to a red node in the logical graph in Fig.~\ref{fig:Keilingtonow}(e). All other $3$-photon clusters in Fig.~\ref{fig:Keilingtonow}(b) will have all their three photons measured in fusion operations and so, all these $3$-photon clusters are represented as black nodes in the logical graph. In the logical graph, a node represents an $n$-photon cluster, and a node's degree equals the number of its photons that will be measured in fusion attempts (and hence destroyed). A bond in the logical graph represents a fusion attempt, which is successfully activated with probability $\lambda$. With this new interpretation, the modified site-bond percolation discussed above can now be seen as simple single-parameter bond percolation on the logical graph, where each bond is independently activated with probability $\lambda$. It is simple to verify numerically (see the plot in Fig.~\ref{fig:Keilingtonow}(e)) that the bond percolation threshold equals $\lambda_c \approx 0.672$, as expected. 

The black nodes disappear during the fusion attempts but help provide long-range connections. Only the red nodes, which in the example of Fig.~\ref{fig:Keilingtonow}(e) contain a single photon each after the fusion attempts, remain as part of the giant connected component, which is subsequently renormalized for quantum computing. Bond percolation guarantees that if $N$ is the number of nodes in the logical graph $G$, and if $\lambda > \lambda_c$, the bond percolation threshold of $G$, then there is a unique giant connected component (GCC), i.e., a large cluster with $O(N)$ nodes. These $O(N)$ nodes have both red nodes and black nodes. However, it is simple to argue that there are $O(N)$ red nodes in the GCC. 

Finally, note that in the example shown in Fig.~\ref{fig:Keilingtonow}(e), even though the logical graph---which describes how to lay out the microclusters prior to fusion attempts---is a non-planar two-layer graph, the physical giant cluster (of photons) obtained from percolation is a subgraph of the planar square lattice. 

\subsubsection{The diamond lattice and the (10,3)-b logical lattice}\label{sec:diamond10_3b}

\begin{figure}
    \includegraphics[width=\columnwidth]{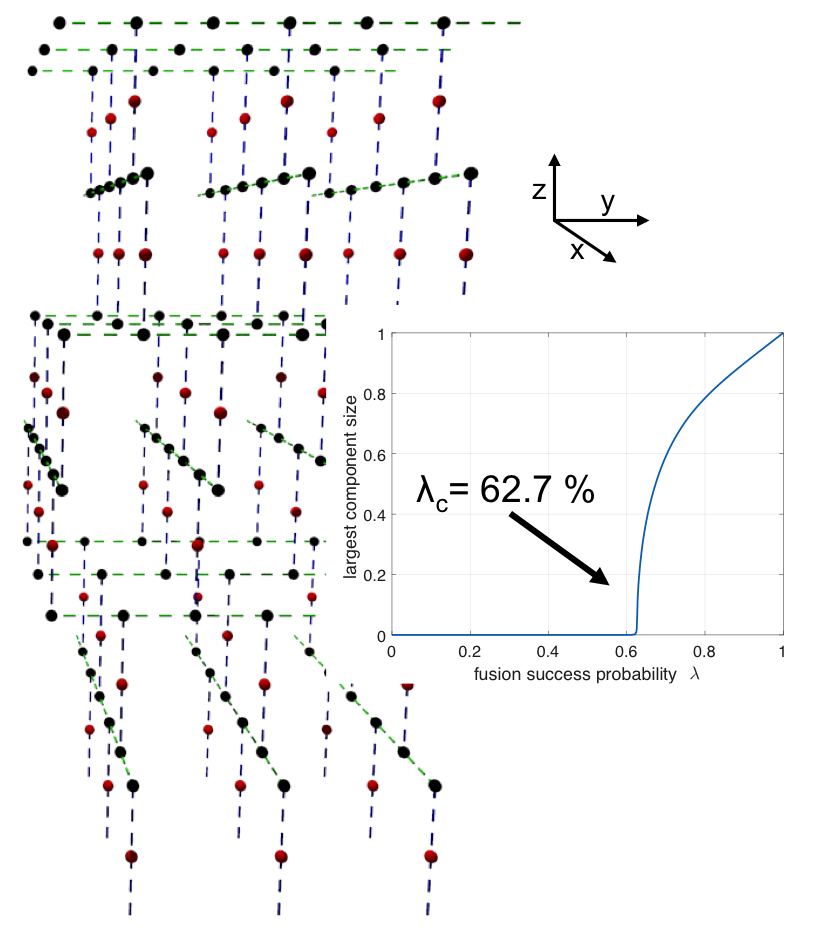}
    \caption{A 3D (10,3)-b lattice modified with additional nodes at the centers of each vertical bond. Pure bond percolation on this logical lattice corresponds to assembling the 3D diamond lattice using $3$-photon microclusters discussed in~\cite{2015.PRL.Gimeno-Segovia-Rudolph.3GHZtoBallisticQC}. Percolation threshold was evaluated by the Newman-Ziff method on a lattice with $\sim 10^6$ bonds.} 
    \label{fig:103b3D}
\end{figure}
If one repeats the steps outlined in Sections~\ref{sec:Kielingsquare},~\ref{sec:Mercedessquare} and~\ref{sec:purebondsquare} for the 3D diamond lattice, i.e., lay out three $3$-photon clusters in vertical arrangements as in Fig.~\ref{fig:Keilingtonow}(b) at each degree-$4$ node of the 3D diamond lattice---laid out in the layered 3D brickwork configuration as shown in~\cite{2015.PRL.Gimeno-Segovia-Rudolph.3GHZtoBallisticQC}---and map it to a logical graph as described above, one obtains the logical lattice shown in Fig.~\ref{fig:103b3D}. This is the (10,3)-b lattice~\cite{2013.JSMTE.Tran-Small.3Dlattdeg3perc} with one extra node inserted at the center of each of the vertical bonds. We refer to this as the `modified" (10,3)-b lattice. The red nodes, as before, correspond to $3$-photon microclusters with one unmeasured photon, whereas the black nodes correspond to $3$-photon microclusters, all of whose photons will be measured in fusion attempts. We evaluated the bond percolation threshold of this modified (10,3)-b lattice using the Newman-Ziff algorithm, and obtained $\lambda_c \approx 0.627$, which agrees with, and sharpens the result of~\cite{2015.PRL.Gimeno-Segovia-Rudolph.3GHZtoBallisticQC} (i.e., $\lambda_c \approx 0.625$); but is now interpreted as a standard bond percolation threshold.

\subsection{General picture for ballistic LOQC}\label{sec:generalpicture}

The discussion in Section~\ref{sec:fusingclustersonlattice} logically leads to a new approach to constructing a large percolated network of photons for ballistic LOQC. We directly pick an $N$-node logical graph $G$, each node of which represents an $n$-photon microcluster. We color the nodes based on how many photons we intend to leave unmeasured, or equivalently, the node's intended degree in the logical graph. If the node's degree is $d$, $1 \le d \le n$, we give it color $n-d$, the number of photons in the microcluster at that node that are left unmeasured. See Fig.~\ref{fig:Keilingtonow}(d). The goal is, given $n$, to pick a logical graph and node coloring such that one gets the lowest possible bond percolation threshold. In addition, for the percolated cluster to be useful for QC, one must be able to argue that: (a) there are $O(N)$ non-zero-color nodes in the GCC, and (b) the resulting physical cluster of photons can be embedded in a universal resource for quantum computation (e.g. a square lattice) for any possible result of the probabilistic fusions.

\begin{figure}
    \includegraphics[width=\columnwidth]{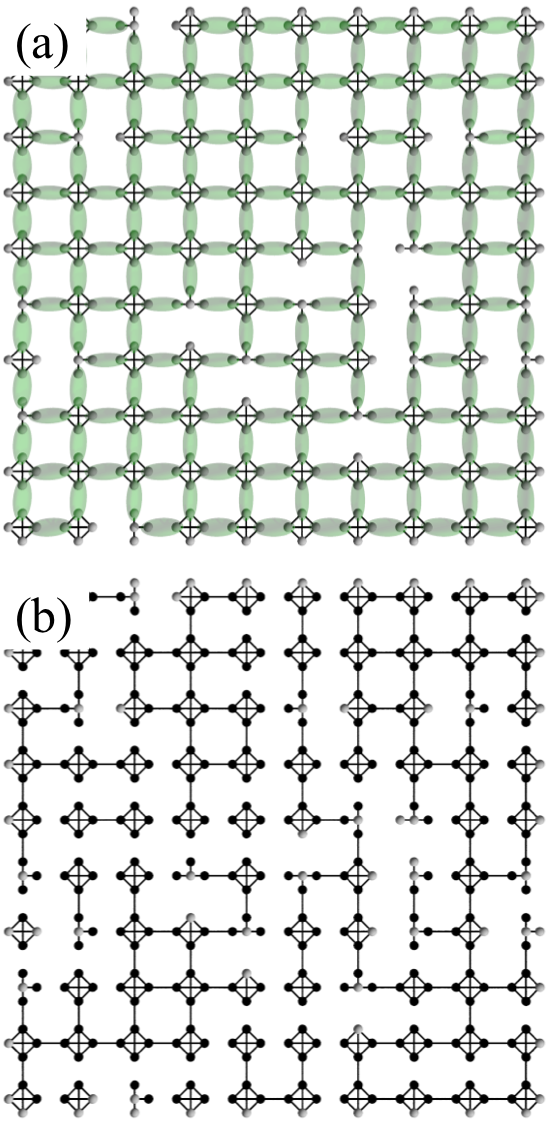}
    \caption{(a) $4$-node microclusters laid out on nodes of a square lattice. A random $\alpha = 0.3$ fraction of microclusters are put in star configuration the central photon of which will not be measured in any fusion operation. All other photons are measured in fusion attempts. (b) A random instance after the fusion attempts, assuming that each fusion succeeds with probability $\lambda = 0.6$. The measured photons are colored black. The unmeasured photons (colored white) in the giant component of the percolated lattice form the backbone random graph that is renormalized into a fully connected 2D topology for universal cluster-state quantum computing.} 
    \label{fig:generalpicture}
\end{figure}
Let us assume $G$ is a regular lattice with uniform node degree $d$. Let us also assume that we have access to $d$-node microclusters. One strategy for selecting nodes in $G$ designated to have non-zero-color is to pick a random fraction, $\alpha$, of the $N$ nodes in $G$ as color-$1$ and populate them with $d$-photon star clusters. Clearly, these nodes will have one less degree ($d-1$). We then populate $d$-photon clique clusters at the remaining $(1-\alpha)N$ degree-$d$ nodes. These nodes have color $0$ and hence all the photons in the cliques will be measured in the fusions. If $\alpha$ is small, then the fusion success probability exceeding the bond percolation threshold of $G$, i.e., $\lambda > p_c(G)$, should suffice to guarantee percolation. This would mean that $\lambda_c^{(n)} \le {\rm min}_{G(V,E): deg(V)=n}\,p_c(G)$. In order to prove this formally, one needs to argue that conditions (a) and (b) in the previous paragraph are met. We leave this for future work. If this conjecture is correct, given that the bond percolation threshold of the degree-$3$ 3D (10,3)-b lattice is $0.546694$~\cite{2013.JSMTE.Tran-Small.3Dlattdeg3perc}, it would mean that $\lambda_c^{(3)} \le 0.546694$ for a 3D lattice. Furthermore, it is possible to generalize the (10,3)-b lattice to higher dimensions, following a procedure similar to the generalization of the ``modified" (10,3)-b lattice in section~\ref{sec:latticeresults}. Under this construction, $\lambda_c \rightarrow 0.5$ as the number of dimensions $\rightarrow \infty$. Combined with our converse of $\lambda_c^{(3)} \ge 0.5$, this would imply that $\lambda_c^{(3)} = 0.5$. We conjecture that a higher dimensional construction with size $n \geq 3$ microclusters can saturate the converse bound which would imply that $\lambda_c^{(n)} = 1/(n-1)$, $\forall n \geq 3$. A schematic of the setup described in the discussion above, with $G$ chosen as the 2D square lattice for illustration, is shown in Fig.~\ref{fig:generalpicture}. 

\section{Fundamental thresholds}\label{sec:fundthresholds}

We begin this section with new results on achievable fusion success thresholds using $3$-photon microclusters in Section~\ref{sec:latticeresults}, i.e., tighter upper bounds on $\lambda_c^{(3)}$ compared to known results. In Section~\ref{sec:intuitiveconverse} we provide an intuitive proof of our general converse bound $\lambda_c^{(n)} \ge 1/(n-1)$, $\forall n \ge 2$. Finally, in Section~\ref{sec:photloss}, we discuss how losses in devices and inline losses affect the fusion thresholds, and discuss its implications for the resource overhead (number of sources and detectors) for ballistic LOQC in the presence of losses.

\subsection{Achievable thresholds}\label{sec:latticeresults}

\begin{figure}
    \includegraphics[width=\columnwidth]{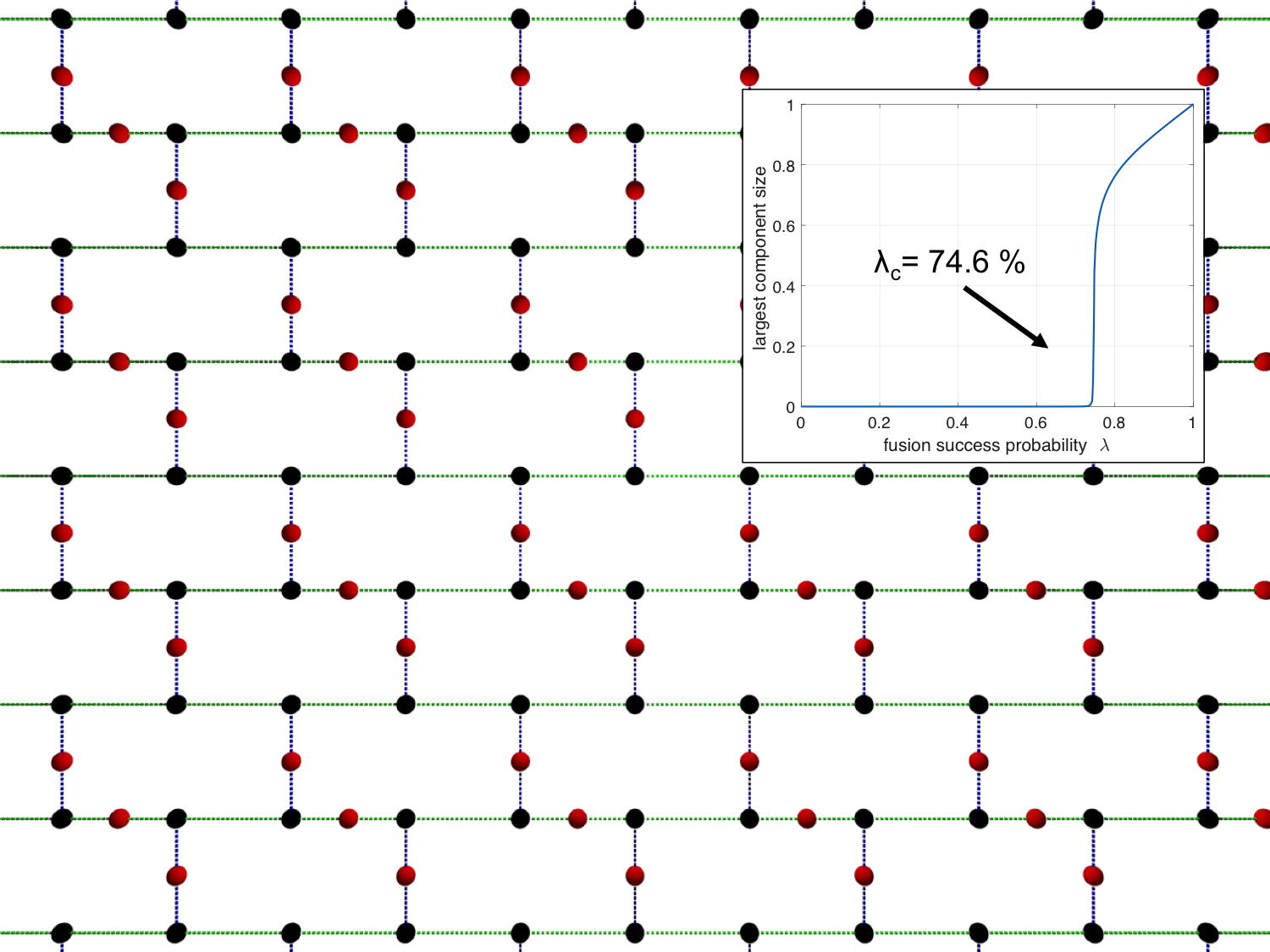}
    \caption{A modified 2D brickwork lattice used as logical graph with node colors as shown yields $\lambda_c \approx 0.746$, which settles an open question in \cite{2016.arXiv.Rudolph.SiPhOptimism} on whether it is possible to attain ballistic LOQC with $3$-photon microclusters with a fully 2D architecture and $\lambda_c < 0.78125$, which is achievable with unentangled ancilla photons. Percolation threshold was evaluated by the Newman-Ziff method on a lattice with $\sim 10^6$ bonds.} 
    \label{fig:2Dmodifiedbrickwork}
\end{figure}
Throughout this section, we take the size of our initial microcluster to be $n = 3$ photons. As in Fig.~\ref{fig:Keilingtonow} (e), blue and green dashed lines correspond to the fusion operations represented by the blue and green ellipses in Fig.~\ref{fig:Keilingtonow}(c), respectively. The degree-$3$ nodes are color-$0$ (black) and hence have $3$-photon clusters all whose photons will be measured in fusion attempts. The degree-$2$ nodes are color-$1$ (red) and have $3$-photon clusters of which only two photons will be measured in fusion attempts.

Let us pick as the logical graph the modified 2D brickwork lattice shown in Fig.~\ref{fig:2Dmodifiedbrickwork}. The bond percolation threshold of this lattice is $\lambda_c \approx 0.746$, as shown in the inset of Fig.~\ref{fig:2Dmodifiedbrickwork}. It is simple to argue that conditions (a) and (b) discussed in Section~\ref{sec:generalpicture} are met, and the resulting percolated cluster is renormalizable. Hence, 
we have shown that even with a 2D lattice, starting with three photon microclusters, it is possible to assemble a resource for universal QC, since $\lambda_c \approx 0.746 < 0.78125$ and two photon fusion with success probability $0.78125$ is achievable with a linear optical circuit boosted with ancilla single photons~\cite{2014.PRL.Ewert-Loock.boostedfusion}. Being able to percolate with a 2D lattice makes ballistic LOQC much easier from the experimental standpoint since a planar integrated photonic waveguide can be used to weave such a cluster. The existence of a 2D lattice with this property was posed as an open question by Rudolph~\cite{2016.arXiv.Rudolph.SiPhOptimism}.

\begin{figure}
    \includegraphics[width=\columnwidth]{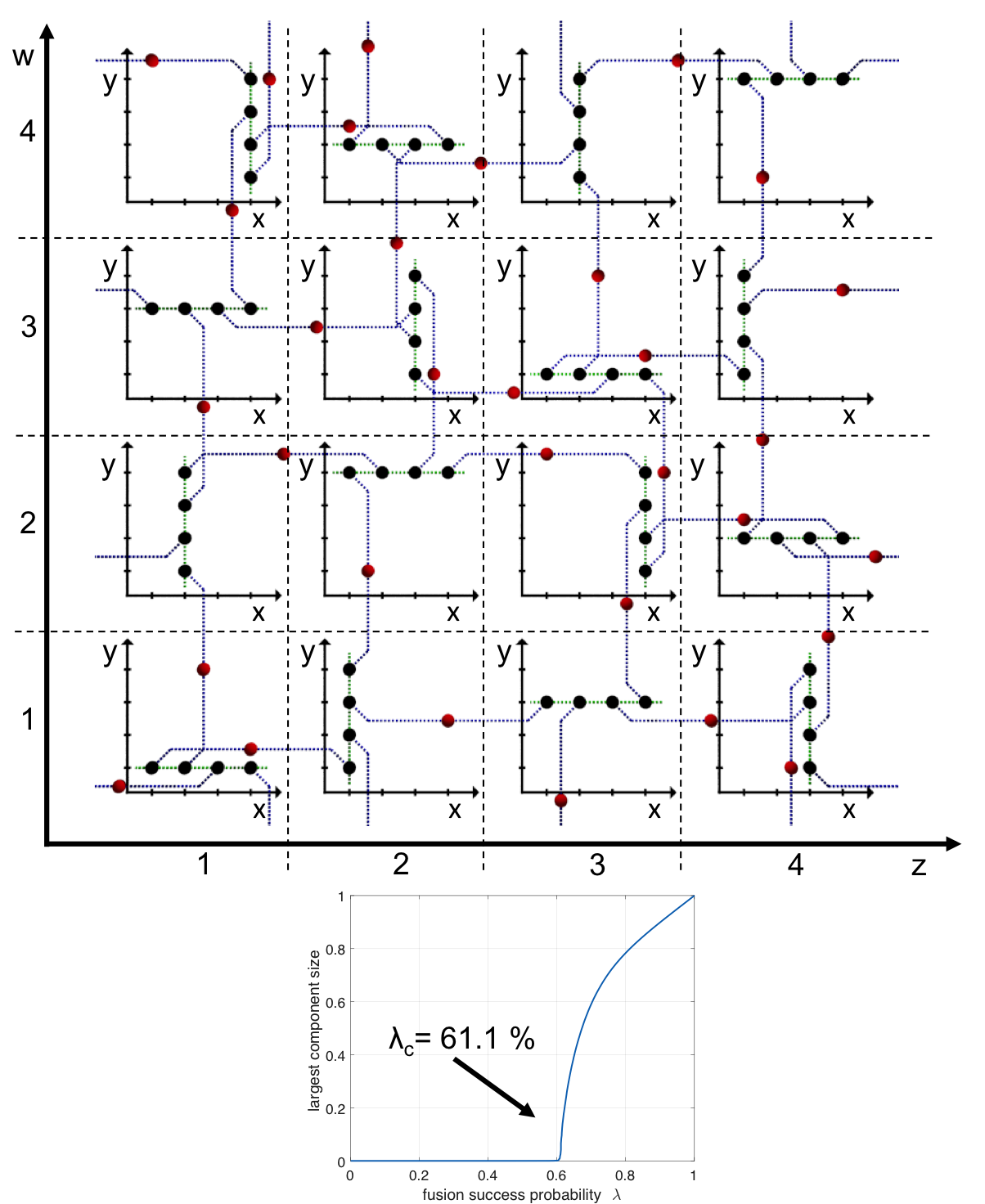}
    \caption{Schematic of the 4D extension of the (10,3)-b lattice, which when used as the logical graph with node colors as shown yields $\lambda_c \approx 0.611$. Percolation threshold was evaluated by the Newman-Ziff method on a lattice with $\sim 10^7$ bonds. The inner plots with $x$ and $y$ axes represent projections of the lattice on the $(x,y)$ plane at the $z$ and $w$ values shown on the outer axes.} 
    \label{fig:103b4Dwithmiddlenode}
\end{figure}

In Section~\ref{sec:diamond10_3b}, we described a logical graph construction of the ``modified" (10,3)-b lattice (Fig.~\ref{fig:103b3D}), using which we reinterpreted the results of~\cite{2015.PRL.Gimeno-Segovia-Rudolph.3GHZtoBallisticQC} as a pure bond percolation threshold, $\lambda_c \approx 0.627$. We now consider a 4D extension of the modified (10,3)-b lattice (Fig.~\ref{fig:103b4Dwithmiddlenode}) as the logical graph. The 3D lattice (Fig.~\ref{fig:103b3D}) comprises $(x,y)$-plane layers of parallel 1D line lattices of black (degree-$3$) nodes stacked along the $z$ direction. The layers alternate between their line lattices pointing in the $x$ and $y$ directions, while neighboring layers are straddled by a layer of red (degree-$2$) nodes. Each black node has two black-node neighbors on either side of the 1D lattice to which it belongs, connected via green bonds, and one red-node neighbor, connected via a blue bond. Along each line lattice of black bonds, the blue bonds alternate between the $+z$ and $-z$ directions. The adjective ``modified" in the name of this lattice refers to the fact that in the standard (10,3)-b lattice, the red nodes are not there, and adjacent $(x,y)$ planes of parallel lattices in alternating directions are directly connected via bonds. Our 4D generalization of the modified (10,3)-b lattice is shown in Fig.~\ref{fig:103b4Dwithmiddlenode}. It consists of a doubly infinite stacking of $(x,y)$-plane layers---of parallel 1D line lattices of black (degree-$3$) nodes---stacked along the $z$ and $w$ directions respectively. Of the three neighboring bonds of a black node, two (green) bonds---connecting to neighboring black nodes in the line lattice to which it belongs---are in the $(x,y)$ plane, whereas one (blue) bond---connecting to a red node which in turns connects via another blue bond to a black node in a neighboring $(x,y)$-plane layer---points in either the $z$ direction or in the $w$ direction. Along each line lattice of black bonds, the blue bonds alternate between directions $+z$, $+w$, $-z$, $-w$, \ldots, and so on. The graph has a period of four in each of the $x$, $y$, $z$ and $w$ dimensions. One period of the lattice is depicted in Fig.~\ref{fig:103b4Dwithmiddlenode}. The inner axes represent an $(x,y)$ plane at a given value of $z$ and $w$. This construction results in longer loops compared to the 3D case while retaining the 3D graph's coordination number (average node degree), which in turn lowers the bond percolation threshold for the 4D logical graph. We find, using a Newman-Ziff simulation performed on a 4D modified (10,3)-b lattice of size $N \sim 10^7$ nodes, that $\lambda_c \approx 0.611$.

\begin{figure}
    \includegraphics[width=0.6\columnwidth]{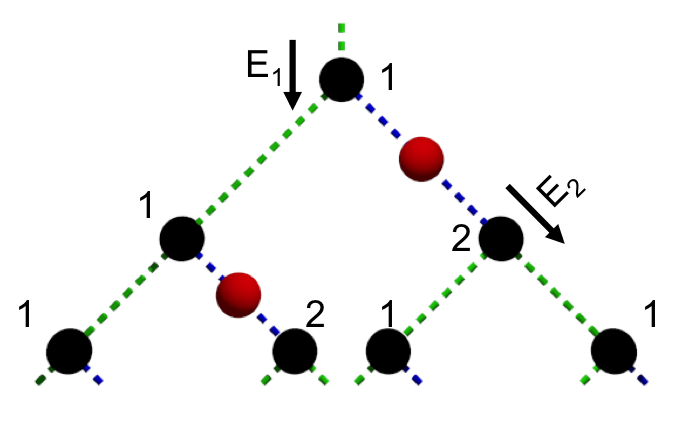}
    \caption{Schematic of the $\infty$-D extension of the (10,3)-b lattice, which when used as logical graph with node colors as shown yields $\lambda_c \approx 0.5898$. Percolation threshold was evaluated analytically.} 
    \label{fig:103binfDwithmiddlenode}
\end{figure}

By adding more dimensions to the aforesaid logical lattice construction, the size of the loops is increased, hence progressively lowering $\lambda_c$. Finally, in the case of the $\infty$-dimensional modified $(10,3)$-b lattice, the loops are infinitely far apart and hence the lattice is locally tree like. The local connectivity of this logical graph is depicted in Fig.~\ref{fig:103binfDwithmiddlenode}. A simple analytical argument, explained below, shows that $\lambda_c \approx 0.5898$ for this limiting construction. This threshold, along with the converse proven in the next section, establishes that $0.5 \le \lambda_c^{(3)} \le 0.5898$, thereby improving upon $\approx 0.625$, the lowest-known fusion probability threshold that is known to be achievable with $3$-photon microclusters~\cite{2015.PRL.Gimeno-Segovia-Rudolph.3GHZtoBallisticQC}. This also is the minimum $\lambda_c^{(3)}$ attainable from higher-dimensional logical lattices of the modified (10,3)-b lattice family. For the entire family of constructions, we argue that conditions (a) and (b) discussed in Section~\ref{sec:generalpicture} are met, and the resulting percolated cluster can be renormalized for QC. 

The locally-tree-like structure of the $\infty$-dimensional modified $(10,3)$-b lattice is shown in Fig.~\ref{fig:103binfDwithmiddlenode}. Similar to the 3D and 4D modified (10,3)-b lattices, each black node has two green bonds and one blue bond (which leads to a black node via a red node and another blue bond). We denote the expected number of children of a node when approached via a green bond as $E_1$ and the expected number of children of a node when approached via a blue bond as $E_2$. When counting the number of children of a node, we only count red nodes since they are the only nodes with unmeasured qubits. Counting children from the top of the Fig~\ref{fig:103binfDwithmiddlenode}, each black node is labelled as $1$ or $2$ depending on the bond from which it is approached. Counting children at the points labelled $E_1$ and $E_2$ yields the equations $E_1 = \lambda E_1 + \lambda + \lambda^2E_2$ and $E_2 = 2\lambda E_1$ where $\lambda$ is the bond probability. For percolation, $E_1 \rightarrow \infty$ and solving the equations with this condition, we find that $\lambda_c+2\lambda_c^3 = 1$, which leads to $\lambda_c = 0.5898$.

A Tree is known not to be a universal resource for QC~\cite{2006.PRL.VanDenNest-Briegel.UnivResMBQC}. However, entangled trees clusters can be used for other applications, e.g., as loss tolerant logical qubits~\cite{2006.PRL.Varnava-Rudolph.CountEC}, with applications to all-photonic quantum repeaters~\cite{2015.NatureComm.Azuma-Lo.AllOptRep, 2017.PRA.Pant-Guha.AllOptRepRescources}. We now show that with a degree-$n$ Bethe Lattice (an infinite tree) as the logical graph, and with $n$-photon microclusters as the initial resource, we can get $\lambda_c^{(n)} = 1/(n-1)$, which saturates the converse bound we prove in the following section. Whether or not $\lambda_c^{(n)} = 1/(n-1)$ can be attained on a lattice whose percolated instance can be renormalized into a logical cluster universal for QC, remains open.

\begin{figure}
    \includegraphics[width=0.6\columnwidth]{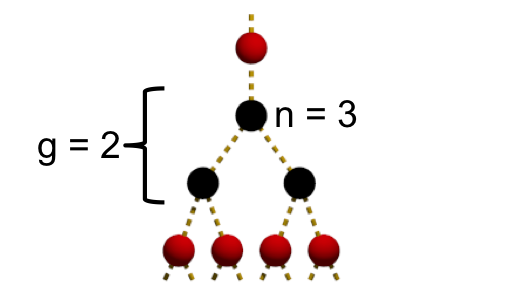}
    \caption{Schematic of the lattice construction used to approach the $\lambda_c = 1/(n-1)$ limit for the case of $n=3$ and $g=2$.} 
    \label{fig:achievability_bethe}
\end{figure}

The logical graph that can be used to approach the $1/(n-1)$ limit is shown in Fig.~\ref{fig:achievability_bethe} for $n = 3$. We start the depiction of our tree at a degree $n-1$ unmeasured node (i.e., a node with an unmeasured qubit), after which there are $g$ generations of degree $n$ black nodes, followed by a generation of unmeasured nodes, followed by $g$ generations of black nodes and so on. In the tree depicted in Fig.~\ref{fig:achievability_bethe}, $g=2$. Starting from an unmeasured node, given a bond probability of $\lambda$, the expected number of unmeasured nodes after $g+1$ generations is $\lambda(n-2)[\lambda(n-1)]^g$. Hence the critical bond percolation probability must satisfy $\lambda_c(n-2)[\lambda_c(n-1)]^g = 1$, which gives us $\lambda_c = (n-2)^{-1/(g+1)}(n-1)^{-g/(g+1)}$. As $g$ increases, we approach the limit of $\lambda_c = 1/(n-1)$. In the argument above, we only count the number of unmeasured nodes and condition (a) of Section~\ref{sec:generalpicture} is satisfied.

In the construction of the Bethe lattice above, the input states are $n$ photon cliques, which are equivalent, up to local operations, to $n$ photon GHZ states. The fusion operation used here (yellow dashed lines), acting on two qubits A and B, consists of a Hadamard gate on qubit A followed by Bell measurement of A and B in the $\left\{ 1/\sqrt{2}(\ket{00} \pm \ket{11}), 1/\sqrt{2}(\ket{01} \pm \ket{10}) \right\}$ basis (also described in \cite{2015.PRA.Zaidi-Rudolph.BallisticLOQC}). Since the order of the Bell measurements is not important, we imagine first applying the fusion operations corresponding to successes. A successful fusion between two cliques removes qubits A and B from the graph and places the rest of the photons in a clique. Hence any two logical nodes that have an edge in Fig.~\ref{fig:achievability_bethe} are part of the same clique and hence connected. A failed fusion results in an $X$ measurement on A and a $Z$ measurement on B. The $Z$ measurement of a qubit simply removes the photon and all its edges. The $X$ measurement of a qubit in a clique has the effect of a $Z$ measurement followed by a Hadamard gate on one of the original neighbors of the qubit. Since a Hadamard gate followed by a Z (resp., X) measurement has the effect of an X (resp., Z) measurement, the result of the failed fusions is simply the removal of the corresponding nodes from the cliques without disturbing the connectivity between any other nodes. Hence the fusion operation described here can be used to create the logical graph in Fig.~\ref{fig:achievability_bethe}.

Finally, as discussed in Section~\ref{sec:generalpicture}, we conjecture that if there is an infinite lattice $G$ of maximum node degree $n$ with bond percolation threshold $p_c$, it is possible to assemble a giant percolated cluster renormalizable for universal QC using $n$-photon microclusters as long as the fusions succeed with probability $\lambda > p_c$. We also conjecture, using an extension of the argument for $n=3$ using an infinite-dimensional modified (10,3)-b lattice, that the converse bound $\lambda_c^{(n)} \ge 1/(n-1)$ is tight for all $n$, i.e., it is possible to construct a logical graph that can be percolated with two-fusion success probability $= 1/(n-1) + \epsilon$, for any $\epsilon > 0$.

\subsection{Converse}\label{sec:intuitiveconverse}

In this section we discuss our converse result: starting with $N$ microclusters each of $n$ photons and using any sequence of two-photon destructive fusion operations, the minimum fusion success probability $\lambda_c$ sufficient to obtain a connected component of $O(N)$ unmeasured photons with high probability is $\geq 1/(n-1)$. A formal proof is provided in the supplemental section. We sketch the intuition behind the proof below.

\begin{figure}
    \includegraphics[width=\columnwidth]{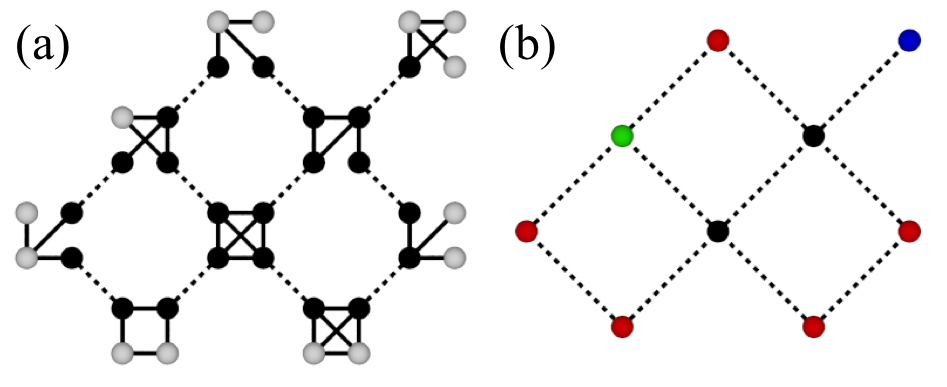}
    \caption{(a) An example of a series of two-node fusions on $n = 4$ sized microclusters. (b) Mapping of the microclusters to nodes in a logical graph. Logical nodes with one, two, three, and four measured physical nodes are colored as Blue, Red, Green, and Black, respectively.} 
    \label{fig:Converse}
\end{figure}

Fig.~\ref{fig:Converse}(a) illustrates an example with $n = 4$ photon microclusters and a set of two-photon fusion attempts shown as dashed lines each of which succeeds with probability $\lambda$, using the graphical interpretation of fusion presented in Fig.~\ref{fig:Keilingtonow}(c). Recalling our convention from Section~\ref{sec:newpicture}, a black photon is one that gets measured in a fusion attempt, hence does not exist after the fusion attempt involving it has happened, regardless of its success. After all the fusions have been attempted, one obtains connected components involving only the white photons. 

Given a large number $N$ of $n$-photon microclusters, our objective is to pick a set of photon pairs on which to attempt fusions (each of which succeeds with probability $\lambda$), such that $\lambda > \lambda_c$ ensures a unique connected component of $O(N)$ white photons with the smallest possible $\lambda_c$. One can argue that the post-fusion connectivity graph that results between the surviving (white) photons, no matter what kind of destructive linear-optical fusion operation is used, can be no more connected than the connectivity between white photons in the graph shown in Fig.~\ref{fig:Converse}(a). In other words, if two white photons have a path connecting them (via black and white photons) in a random instance of the graph in Fig.~\ref{fig:Converse}(a), those two photons would also have a connected path in the actual post-fusion connectivity graph assuming the same success-failure fusion instances, if any linear optical circuit for fusion is employed. 

Fig.~\ref{fig:Converse}(b) shows the mapping of Fig.~\ref{fig:Converse}(a) to a logical graph where each microcluster is replaced by a logical node, similar to Fig.~\ref{fig:Keilingtonow}(d). Here, dashed lines represent bonds in the logical graph that exist with probability $\lambda$. Logical nodes corresponding to microclusters with one, two, three, and four measured (black) photons are colored Blue, Red, Green and Black, respectively. Since the microclusters in Fig.~\ref{fig:Converse}(a) have $n=4$ photons and each photon is associated with at most one fusion attempt, the maximum degree of each logical node in Fig.~\ref{fig:Converse}(b) is $n = 4$. Hence, the post-fusion instance of the logical graph in Fig.~\ref{fig:Converse}(b) represents an instance of bond percolation on some graph of degree four. In general, starting with $n$ photon microclusters, and any sequence of fusion attempts, the resulting instance of the logical graph is a bond percolation instance (with bond success probability of $\lambda$) on some graph of maximum degree $n$. Of all infinite graphs of maximum degree $n$, the minimum bond percolation threshold is that of the degree-$n$ Bethe Lattice, and equals $1/(n-1)$~\cite{1957.MathProcCambPhilSoc.Broadbent-Hammersley.Percolation, 1983.JPAMG.Sahimi-Davis.BPthresh}. Finally, since each of the $N$ logical nodes in Fig.~\ref{fig:Converse}(b) maps to $n$ photons in Fig.~\ref{fig:Converse}(a), with $n$ finite, the absence of a connected component with $O(N)$ logical nodes in the post-fusion instance of the logical lattice of Fig.~\ref{fig:Converse}(b) implies the absence of a connected component of $O(N)$ white photons in the post-fusion instance of the physical photonic cluster shown in Fig.~\ref{fig:Converse}(a). This completes our proof that $\lambda_c^{(n)} \ge 1/(n-1)$. Since the Bethe Lattice is a tree, it cannot be renormalized into a logical cluster state that is needed for universal quantum computing~\cite{2006.PRL.VanDenNest-Briegel.UnivResMBQC}. Therefore, the above proof does not establish the tightness of the aforesaid bound. As explained however in the context of $n=3$ in Section~\ref{sec:newpicture}, we believe that the bound $\lambda_c^{(n)} \ge 1/(n-1)$ is tight, and is achievable by going to progressively higher dimensional equivalents of the (10,3)-b lattice for the logical graph, since the logical graph's local topology increasingly resembles that of the Bethe Lattice while retaining the renormalizability of its bond-percolated instance for any finite dimension. But a rigorous proof of the above and a fully-specified construction of the achievability of the $1/(n-1)$ threshold for universal QC is beyond the scope of this paper.

Using similar reasoning, it is also possible to show that starting with $N$ microclusters of size $n$ and using any sequence of $m$-node (destructive) fusion operations, the fusion success probability threshold required to obtain a component of $O(N)$ unmeasured photons satisfies $\lambda_c^{(n,m)} \geq 1/\left[(n-1)(m-1)\right]$. Very little is known about linear-optical fusion of more than two photons at once and their associated success probabilities. Therefore, it is unclear whether the above lower bound on $\lambda_c^{(n,m)}$ is tight.

\subsection{Photon Loss}\label{sec:photloss}

In this section, we present a lower bound on the percolation threshold $\lambda_c^{(n)}$ in the presence of photon loss. Our results suggest that in the presence of loss, there may be an optimum size of the input microclusters that achieves the lowest fusion success probability necessary for achieving percolation, and hence allows for the greatest tolerance to photon loss. We use a loss model inspired by a recently proposed method to produce photonic microclusters using quantum dot emitters~\cite{2009.PRL.Lindner-Rudolph.ClusMachGun, 2016.Science.Schwartz-Gershoni.QDClusGen}. In this method, a quantum dot-confined electron is replaced by a confined dark exciton and this dark exciton subsequently interacts with a series of single photons that are initially unentangled. The creation of an $n$ photon GHZ state (a microcluster in the line lattice graph state) involves $n$ entangling operations. We assume that each photon produced by the source experiences the same transmissivity $\eta_0^n$ with $\eta_0 < 1$, and that detector and waveguide losses are lumped into the parameter $\eta_0$. The rationale behind this stems from the assumption that the exciton loss acts independently on each photon and that the entire microcluster needs to be produced at the same time: the transmissivity experienced by the $k^{\rm{th}}$ photon, $\eta_k =\eta_{\rm{exciton}}^k \eta_{\rm{waveguide}}^{n-k} = \eta_0^n r^k$, with $\eta_0 = {\rm{max}}(\eta_{\rm{exciton}}, \eta_{\rm{waveguide}})$ and $r = {\rm{min}}(\eta_{\rm{exciton}}, \eta_{\rm{waveguide}})/{\rm max}(\eta_{\rm{exciton}}, \eta_{\rm{waveguide}} )$. Since $r\leq1$, $\eta_k \le \eta_0^n$, $\forall k = 1, \ldots, n$. Therefore assuming that each photon in the $n$-photon microcluster experiences identical transmissivity $\eta_0^n$ is an optimistic model which leads to a higher inferred graph connectivity in the post-fusion cluster compared to the true connectivity. Since we are seeking a lower bound on $\lambda_c$, this is acceptable.


In the absence of any photon loss, starting with $N$ entangled microclusters of $n$ photons each, the minimum value of two-photon fusion success probability {\em necessary} to obtain an $O(N)$ photon connected component satisfies the lower bound $\lambda_c \ge 1/(n-1)$, which if our conjecture explained above is true is also {\em sufficient} (achievable) for percolation. In the presence of photon loss, the above lower bound on $\lambda_c$ remains a valid, yet trivial, lower bound. We would like a non-trivial lower bound on $\lambda_c$ that is a function of $\eta_0$ and $n$, such that the lower bound increases with decreasing $\eta_0$.

\begin{figure}
\includegraphics[width=\columnwidth]{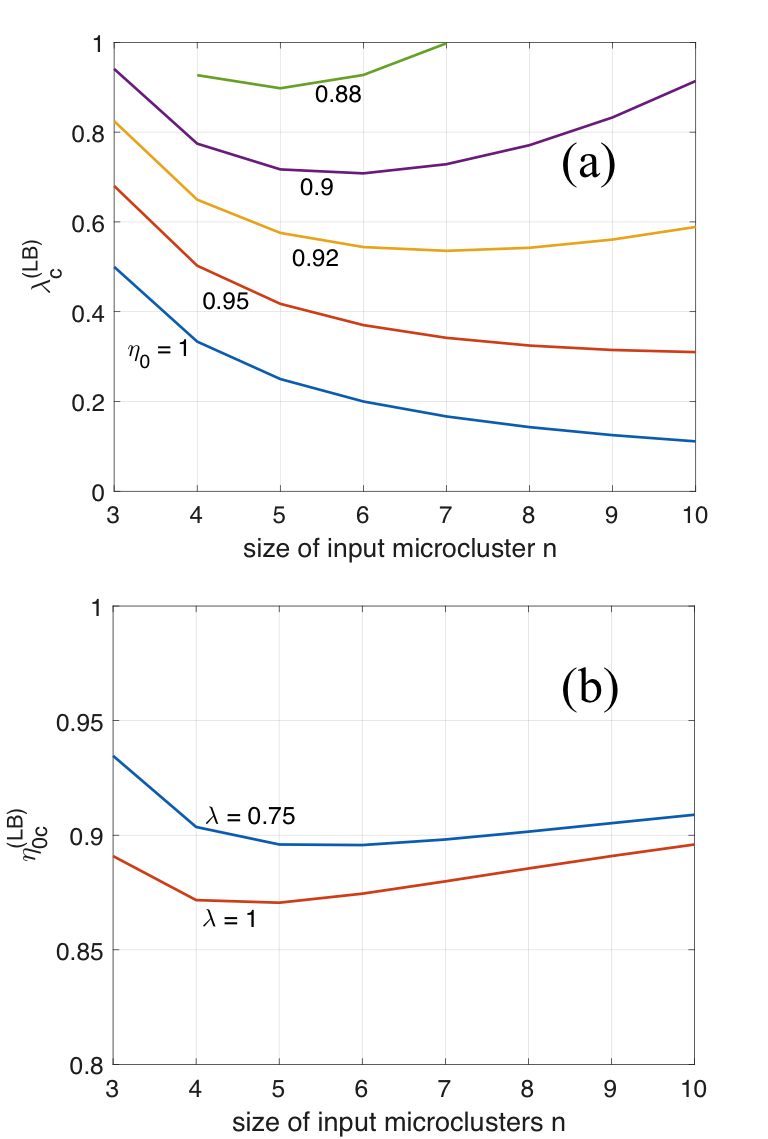}
\caption{(a) A loss-dependent lower bound $\lambda_c^{\rm{(LB)}}$ on the critical fusion probability $\lambda_c$ as a function of the input microcluster size $n$ for different values of $\eta_0$; (b) a loss-dependent lower bound $\eta_{0c}^{\rm{(LB)}}$ on the critical loss parameter $\eta_{0}$ as a function of $n$ for different values of fusion success probability $\lambda$.} 
	\label{fig:phloss}
\end{figure}

Let us say the success probability of a two-photon fusion operation is $\lambda$. As discussed above, there are two types of photons, ones that are measured in fusion attempts and ones that are not. The latter type of photons constitute the renormalizable percolated giant component when $\lambda > \lambda_c$. In the presence of losses, both types of photons undergo loss. Loss of a photon that was measured in a (destructive) fusion attempt is {\em detected}, since the number of expected detector clicks at the output of the linear-optical circuit for fusion is lower than that is expected. On the other hand, the loss of the unmeasured photons cannot be detected (assuming we do not have access to a quantum non-demolition measurement). This results in the post-fusion cluster to be in a mixed state, a probabilistic mixture of all possible combinations of the unmeasured photons being lost or not. It is not known whether such a mixed state cluster (i.e., without the knowledge of which of the unm easured photons were lost)---even if percolated---can be renormalized into a logical lattice or not, unless each photon (qubit) in the model considered in this paper is replaced by a loss-protected logical qubit, e.g., a tree qubit~\cite{2006.PRL.Varnava-Rudolph.CountEC}. However, since we are seeking a lower bound on $\lambda_c$, we only need to consider a pure graph state that is more connected that the true post-fusion cluster. The simplest way to do so is to pick the post-fusion cluster state where none of the unmeasured photons were lost. With these assumptions, each fusion (a bond in the logical graph) succeeds with probability $\lambda ,\eta_0^{2n}$. Therefore, following the arguments in Section~\ref{sec:intuitiveconverse}, if $\lambda < 1/\left[(n-1)\eta_0^{2n}\right]$, the post-fusion cluster state cannot have a connected component with $O(N)$ unmeasured photons. Hence, we have the following loss-dependent lower bound: $\lambda_c \ge 1/\left[(n-1)\eta_0^{2n}\right] \equiv \lambda_c^{\rm{(LB)}}$. This lower bound is plotted in Fig.~\ref{fig:phloss}(a) for different values of $\eta_0$. We find that while increasing the size of the input microclusters $n$ in the lossless case ($\eta_0 = 1$) always results in a reduction in the necessary fusion success probability for percolation, in the presence of finite losses ($\eta_0 < 1$), there is an optimum value of $n$ that gives the minimum fusion probability. For example, for $\eta_0 = 0.9$, $n = 6$ sized microclusters yield the lowest necessary fusion success probability threshold for percolation.

Conversely, for a given fusion success probability $\lambda$, there exists a threshold $\eta_{0c}$, s.t., if $\eta_0 < \eta_{0c}$, the post-fusion cluster cannot be percolated. We thus have a lower bound $\eta_{0c} \ge \eta_{0c}^{\rm{(LB)}}$, where $\eta_{0c}^{\rm{(LB)}} = \left[1/{\lambda(n-1)}\right]^{1/(2n)}$. In Fig.~\ref{fig:phloss}(b), we plot $\eta_{0c}^{\rm (LB)}$ for different values of $n$, for $\lambda = 0.75$ and $\lambda = 1$. There is an optimum value of $n$ which gives the best loss tolerance, e.g., for $\lambda = 0.75$, six photon microclusters gives the best loss tolerance of $\eta_{0c}^{\rm{(LB)}} = 0.8957$ which corresponds to a loss of $48.36\%$ seen by each photon. Furthermore, we find that going from $\lambda = 0.75$, which is attainable using four single ancilla photons and (lossless) linear optics~\cite{2014.PRL.Ewert-Loock.boostedfusion} to deterministic fusion ($\lambda = 1$), $\eta_{0c}^{\rm{(LB)}}$ only decreases slightly, i.e., the equivalent per-photon loss threshold increases from $0.896$ to $0.871$. Hence, when losses are accounted for in ballistic cluster state creation, the advantage in having a fully deterministic fusion may be relatively small.

Finally, it may be possible to get a tighter lower bound on $\lambda_c$ by using a more sophisticated loss model. For example, the assumption that the exciton loss acts independently on each photon is not entirely accurate, resulting in positively-correlated bonds within the microclusters. Further, the assumption we made about all unmeasured photon not being lost may affect the tightness of the lower bound. However, this last assumption may not have been that ominous, considering our conjectured tightness of the lower bound $\lambda_c \ge \lambda_c^{\rm{(LB)}} = 1/(n-1)$ in the lossless case ($\eta_0 = 1$) was based on a construction where the fraction of photons $\alpha$ in the logical graph that are left unmeasured goes towards zero.

What we leave unaddressed in this paper, are constructive solutions for ballistic photonic quantum computing, or achievability results (i.e., upper bounds on $\lambda_c$ for a given $\eta_0$ or upper bonds on $\eta_{0c}$ for a given $\lambda$) in the presence of photon loss. This will require one to determine how to construct a loss-error-corrected logical lattice fully ballistically (perhaps using tree error correction but with randomly-grown trees) the percolated instance of which can be provably renormalized into a logical cluster state, every node of which is an appropriately loss-protected photonic qubit. In recent work~\cite{2015.PRL.Gimeno-Segovia-Rudolph.3GHZtoBallisticQC, 2015.PhDThesis.Mercedes}, Gimeno-Segovia estimated loss tolerance of ballistic creation of certain percolated lattices by a strategy where one measures all the neighbors of lost photons in the $Z$ basis. This method also accounts for losses in the photons undergoing fusion operations but not in the photons that remain unmeasured, thereby also not proving achievability results in the presence of photon loss. This is an important question that must be addressed systematically not only for photon loss, but for other forms of qubit error models specific to linear-optical qubits such as mode mismatch and detector dark clicks, for this scheme to become a practically feasible solution to scalable quantum computing.

\section{Conclusions and open problems}\label{sec:conclusions}

In this paper, we analyzed fundamental thresholds on the success probability of two-photon linear optical fusion operations for preparing large renormalizable photonic clusters for universal cluster-model quantum computing. We introduced a new percolation framework to study this problem, based on which we developed new constructions with improved thresholds and geometric properties over known results, and found fundamental bounds on the thresholds. We also discussed how losses---inline losses and losses at sources and detectors---affect the bounds on the percolation threshold, using loss models inspired from a recently-proposed method to produce photonic microclusters using quantum dots.

Many interesting open questions remain. One major fundamental open question is the minimum overhead required (i.e., number of physical photons in a logical qubit) to error correct for a given amount of loss rate on each physical photon. There has been considerable research to date on quantum error correction to tackle optical loss. This includes the work on tree codes~\cite{2006.PRL.Varnava-Rudolph.CountEC, 2007.NJP.Varnava-Rudolph.CountFactECMemory, 2008.PRL.Varnava-Rudolph.PhotonLoss-LOQCscaling} and surface codes ~\cite{2010.PRL.Barrett-Stace.TopolPhotonLoss}. The percolation approach discussed in this paper is another way to code for optical loss, but as discussed in the previous section, more work needs to be done to design fully ballistic (feedback-free) constructions for renormalizing an error-free logical cluster for quantum computing. One way to do this would be to replace each physical photon in the construction discussed in this paper by loss-protected photonic qubits, e.g., using tree clusters.

Furthermore, a big practical challenge in making scalable photonic quantum computing feasible is to develop error correction techniques to correct other (non loss) errors, the two most important being mode-mismatch errors and multi-photon events---both in the sources as well as in the detectors (i.e., dark clicks). The whole construction described in this paper relies on a perfect interferometer processing many identical photons, akin to a giant Hong-Ou-Mandel interferometer~\cite{2015.Science.Carolan-Laing.UnivLO, 2015.PRA.Mower-Englund.QPPTheory, 2015.arXiv.Harris-Englund.BosTransQPP}. Mode mismatch can be caused due to the interfering photons not being in identical modes, or small errors and deviations in the splitting ratios of beamsplitters and phase errors. Our paper reinforces the message from the recent work of Rudolph and colleagues, that sources that can directly generate entangled clusters of a small number of photons would be a very valuable resource, and developing new ideas and designs of such photonic sources would be an extremely worthwhile pursuit.

\begin{acknowledgments}
MP, DE, and SG acknowledge support from the DARPA seedling project {\em Scalable Engineering of Quantum Optical Information Processing Architectures} (SEQUOIA), under US Army contract number W31P4Q-15-C-0045. MP and DE acknowledge support from the Air Force Office of Scientific Research MURI (FA9550-14-1-0052). SG acknowledges support from the Office of Naval Research MURI on Optical Computing under US Navy contract number N00014-16-C-2069. MP, DE and SG acknowledge extremely valuable discussions with Terry Rudolph, Pete Shadbolt and Mercedes Gimeno-Segovia during their visit to Boston in April 2016 which was partially supported by the MIT-Imperial College London Seed Fund. 
\end{acknowledgments}

\bibliography{LOQCpercolationpaper}

\begin{thebibliography}{37}%
\makeatletter
\providecommand \@ifxundefined [1]{%
 \@ifx{#1\undefined}
}%
\providecommand \@ifnum [1]{%
 \ifnum #1\expandafter \@firstoftwo
 \else \expandafter \@secondoftwo
 \fi
}%
\providecommand \@ifx [1]{%
 \ifx #1\expandafter \@firstoftwo
 \else \expandafter \@secondoftwo
 \fi
}%
\providecommand \natexlab [1]{#1}%
\providecommand \enquote  [1]{``#1''}%
\providecommand \bibnamefont  [1]{#1}%
\providecommand \bibfnamefont [1]{#1}%
\providecommand \citenamefont [1]{#1}%
\providecommand \href@noop [0]{\@secondoftwo}%
\providecommand \href [0]{\begingroup \@sanitize@url \@href}%
\providecommand \@href[1]{\@@startlink{#1}\@@href}%
\providecommand \@@href[1]{\endgroup#1\@@endlink}%
\providecommand \@sanitize@url [0]{\catcode `\\12\catcode `\$12\catcode
  `\&12\catcode `\#12\catcode `\^12\catcode `\_12\catcode `\%12\relax}%
\providecommand \@@startlink[1]{}%
\providecommand \@@endlink[0]{}%
\providecommand \url  [0]{\begingroup\@sanitize@url \@url }%
\providecommand \@url [1]{\endgroup\@href {#1}{\urlprefix }}%
\providecommand \urlprefix  [0]{URL }%
\providecommand \Eprint [0]{\href }%
\providecommand \doibase [0]{http://dx.doi.org/}%
\providecommand \selectlanguage [0]{\@gobble}%
\providecommand \bibinfo  [0]{\@secondoftwo}%
\providecommand \bibfield  [0]{\@secondoftwo}%
\providecommand \translation [1]{[#1]}%
\providecommand \BibitemOpen [0]{}%
\providecommand \bibitemStop [0]{}%
\providecommand \bibitemNoStop [0]{.\EOS\space}%
\providecommand \EOS [0]{\spacefactor3000\relax}%
\providecommand \BibitemShut  [1]{\csname bibitem#1\endcsname}%
\let\auto@bib@innerbib\@empty
\bibitem [{\citenamefont {Knill}\ \emph {et~al.}(2001)\citenamefont {Knill},
  \citenamefont {Laflamme},\ and\ \citenamefont
  {Milburn}}]{2001.Nature.Knill-Milburn.KLM}%
  \BibitemOpen
  \bibfield  {author} {\bibinfo {author} {\bibfnamefont {E.}~\bibnamefont
  {Knill}}, \bibinfo {author} {\bibfnamefont {R.}~\bibnamefont {Laflamme}}, \
  and\ \bibinfo {author} {\bibfnamefont {G.~J.}\ \bibnamefont {Milburn}},\
  }\href {\doibase 10.1038/35051009} {\bibfield  {journal} {\bibinfo  {journal}
  {Nature}\ }\textbf {\bibinfo {volume} {409}},\ \bibinfo {pages} {46}
  (\bibinfo {year} {2001})}\BibitemShut {NoStop}%
\bibitem [{\citenamefont {Raussendorf}\ and\ \citenamefont
  {Briegel}(2001)}]{2001.PRL.Raussendorf-Briegel.ClusterStateComp}%
  \BibitemOpen
  \bibfield  {author} {\bibinfo {author} {\bibfnamefont {R.}~\bibnamefont
  {Raussendorf}}\ and\ \bibinfo {author} {\bibfnamefont {H.~J.}\ \bibnamefont
  {Briegel}},\ }\href {\doibase 10.1103/PhysRevLett.86.5188} {\bibfield
  {journal} {\bibinfo  {journal} {Physical Review Letters}\ }\textbf {\bibinfo
  {volume} {86}},\ \bibinfo {pages} {5188} (\bibinfo {year}
  {2001})}\BibitemShut {NoStop}%
\bibitem [{\citenamefont {Browne}\ and\ \citenamefont
  {Rudolph}(2005)}]{2005.PRL.Browne-Rudolph.Clusterstates}%
  \BibitemOpen
  \bibfield  {author} {\bibinfo {author} {\bibfnamefont {D.~E.}\ \bibnamefont
  {Browne}}\ and\ \bibinfo {author} {\bibfnamefont {T.}~\bibnamefont
  {Rudolph}},\ }\href {\doibase 10.1103/PhysRevLett.95.010501} {\bibfield
  {journal} {\bibinfo  {journal} {Physical Review Letters}\ }\textbf {\bibinfo
  {volume} {95}},\ \bibinfo {pages} {010501} (\bibinfo {year}
  {2005})}\BibitemShut {NoStop}%
\bibitem [{\citenamefont {Kieling}\ \emph {et~al.}(2007)\citenamefont
  {Kieling}, \citenamefont {Rudolph},\ and\ \citenamefont
  {Eisert}}]{2007.PRL.Kieling-Eisert.PercolationQC}%
  \BibitemOpen
  \bibfield  {author} {\bibinfo {author} {\bibfnamefont {K.}~\bibnamefont
  {Kieling}}, \bibinfo {author} {\bibfnamefont {T.}~\bibnamefont {Rudolph}}, \
  and\ \bibinfo {author} {\bibfnamefont {J.}~\bibnamefont {Eisert}},\ }\href
  {\doibase 10.1103/PhysRevLett.99.130501} {\bibfield  {journal} {\bibinfo
  {journal} {Physical Review Letters}\ }\textbf {\bibinfo {volume} {99}},\
  \bibinfo {pages} {130501} (\bibinfo {year} {2007})}\BibitemShut {NoStop}%
\bibitem [{\citenamefont {Kieling}\ and\ \citenamefont
  {Eisert}(2009)}]{2009.Book.Kieling-Eisert.Percolation}%
  \BibitemOpen
  \bibfield  {author} {\bibinfo {author} {\bibfnamefont {K.}~\bibnamefont
  {Kieling}}\ and\ \bibinfo {author} {\bibfnamefont {J.}~\bibnamefont
  {Eisert}},\ }\href {\doibase 10.1007/978-3-540-85428-9_10} {\bibfield
  {journal} {\bibinfo  {journal} {Lecture Notes in Physics}\ }\textbf {\bibinfo
  {volume} {762}},\ \bibinfo {pages} {287} (\bibinfo {year} {2009})},\ \Eprint
  {http://arxiv.org/abs/0712.1836} {arXiv:0712.1836} \BibitemShut {NoStop}%
\bibitem [{\citenamefont {Gimeno-Segovia}\ \emph {et~al.}(2015)\citenamefont
  {Gimeno-Segovia}, \citenamefont {Shadbolt}, \citenamefont {Browne},\ and\
  \citenamefont {Rudolph}}]{2015.PRL.Gimeno-Segovia-Rudolph.3GHZtoBallisticQC}%
  \BibitemOpen
  \bibfield  {author} {\bibinfo {author} {\bibfnamefont {M.}~\bibnamefont
  {Gimeno-Segovia}}, \bibinfo {author} {\bibfnamefont {P.}~\bibnamefont
  {Shadbolt}}, \bibinfo {author} {\bibfnamefont {D.~E.}\ \bibnamefont
  {Browne}}, \ and\ \bibinfo {author} {\bibfnamefont {T.}~\bibnamefont
  {Rudolph}},\ }\href {\doibase 10.1103/PhysRevLett.115.020502} {\bibfield
  {journal} {\bibinfo  {journal} {Physical Review Letters}\ }\textbf {\bibinfo
  {volume} {115}},\ \bibinfo {pages} {020502} (\bibinfo {year}
  {2015})}\BibitemShut {NoStop}%
\bibitem [{\citenamefont {Zaidi}\ \emph {et~al.}(2015)\citenamefont {Zaidi},
  \citenamefont {Dawson}, \citenamefont {{Van Loock}},\ and\ \citenamefont
  {Rudolph}}]{2015.PRA.Zaidi-Rudolph.BallisticLOQC}%
  \BibitemOpen
  \bibfield  {author} {\bibinfo {author} {\bibfnamefont {H.~A.}\ \bibnamefont
  {Zaidi}}, \bibinfo {author} {\bibfnamefont {C.}~\bibnamefont {Dawson}},
  \bibinfo {author} {\bibfnamefont {P.}~\bibnamefont {{Van Loock}}}, \ and\
  \bibinfo {author} {\bibfnamefont {T.}~\bibnamefont {Rudolph}},\ }\href
  {\doibase 10.1103/PhysRevA.91.042301} {\bibfield  {journal} {\bibinfo
  {journal} {Physical Review A - Atomic, Molecular, and Optical Physics}\
  }\textbf {\bibinfo {volume} {91}},\ \bibinfo {pages} {042301} (\bibinfo
  {year} {2015})},\ \Eprint {http://arxiv.org/abs/1410.3753} {arXiv:1410.3753}
  \BibitemShut {NoStop}%
\bibitem [{\citenamefont {Reck}\ and\ \citenamefont
  {Zeilinger}(1994)}]{1994.PRL.Reck-Zeilinger.DiscreteUnitaryopfrombeamsplitter}%
  \BibitemOpen
  \bibfield  {author} {\bibinfo {author} {\bibfnamefont {M.}~\bibnamefont
  {Reck}}\ and\ \bibinfo {author} {\bibfnamefont {A.}~\bibnamefont
  {Zeilinger}},\ }\href {\doibase 10.1103/PhysRevLett.73.58} {\bibfield
  {journal} {\bibinfo  {journal} {Physical Review Letters}\ }\textbf {\bibinfo
  {volume} {73}},\ \bibinfo {pages} {58} (\bibinfo {year} {1994})}\BibitemShut
  {NoStop}%
\bibitem [{\citenamefont {Grice}(2011)}]{2011.PRL.Grice.BoostedBM}%
  \BibitemOpen
  \bibfield  {author} {\bibinfo {author} {\bibfnamefont {W.~P.}\ \bibnamefont
  {Grice}},\ }\href {\doibase 10.1103/PhysRevA.84.042331} {\bibfield  {journal}
  {\bibinfo  {journal} {Physical Review A}\ }\textbf {\bibinfo {volume} {84}},\
  \bibinfo {pages} {042331} (\bibinfo {year} {2011})}\BibitemShut {NoStop}%
\bibitem [{\citenamefont {Ewert}\ and\ \citenamefont {van
  Loock}(2014)}]{2014.PRL.Ewert-Loock.boostedfusion}%
  \BibitemOpen
  \bibfield  {author} {\bibinfo {author} {\bibfnamefont {F.}~\bibnamefont
  {Ewert}}\ and\ \bibinfo {author} {\bibfnamefont {P.}~\bibnamefont {van
  Loock}},\ }\href {\doibase 10.1103/PhysRevLett.113.140403} {\bibfield
  {journal} {\bibinfo  {journal} {Physical Review Letters}\ }\textbf {\bibinfo
  {volume} {113}},\ \bibinfo {pages} {140403} (\bibinfo {year}
  {2014})}\BibitemShut {NoStop}%
\bibitem [{\citenamefont {Aaronson}\ and\ \citenamefont
  {Arkhipov}(2011)}]{2011.ProcACM.Aaronson-Arkhipov.linear_optics_complexity}%
  \BibitemOpen
  \bibfield  {author} {\bibinfo {author} {\bibfnamefont {S.}~\bibnamefont
  {Aaronson}}\ and\ \bibinfo {author} {\bibfnamefont {A.}~\bibnamefont
  {Arkhipov}},\ }in\ \href {\doibase 10.1145/1993636.1993682} {\emph {\bibinfo
  {booktitle} {Proceedings of the 43rd annual ACM symposium on Theory of
  computing - STOC '11}}}\ (\bibinfo  {publisher} {ACM Press},\ \bibinfo
  {address} {New York, New York, USA},\ \bibinfo {year} {2011})\ p.\ \bibinfo
  {pages} {333}\BibitemShut {NoStop}%
\bibitem [{\citenamefont {Aaronson}\ and\ \citenamefont
  {Arkhipov}(2013)}]{2013.TheoryofComputing.Aaronson-Arkhipov.linear_optics_complexity}%
  \BibitemOpen
  \bibfield  {author} {\bibinfo {author} {\bibfnamefont {S.}~\bibnamefont
  {Aaronson}}\ and\ \bibinfo {author} {\bibfnamefont {A.}~\bibnamefont
  {Arkhipov}},\ }\href {http://www.theoryofcomputing.org/articles/v009a004/}
  {\bibfield  {journal} {\bibinfo  {journal} {Theory of Computing}\ }\textbf
  {\bibinfo {volume} {9}},\ \bibinfo {pages} {143} (\bibinfo {year}
  {2013})}\BibitemShut {NoStop}%
\bibitem [{\citenamefont {Pan}\ and\ \citenamefont
  {Zeilinger}(1998)}]{1998.PRA.Pan-Zeilinger.GHZanalyzer}%
  \BibitemOpen
  \bibfield  {author} {\bibinfo {author} {\bibfnamefont {J.-w.}\ \bibnamefont
  {Pan}}\ and\ \bibinfo {author} {\bibfnamefont {A.}~\bibnamefont
  {Zeilinger}},\ }\href {\doibase 10.1103/PhysRevA.57.2208} {\bibfield
  {journal} {\bibinfo  {journal} {Physical Review A}\ }\textbf {\bibinfo
  {volume} {57}},\ \bibinfo {pages} {2208} (\bibinfo {year}
  {1998})}\BibitemShut {NoStop}%
\bibitem [{\citenamefont {Rudolph}(2016)}]{2016.arXiv.Rudolph.SiPhOptimism}%
  \BibitemOpen
  \bibfield  {author} {\bibinfo {author} {\bibfnamefont {T.}~\bibnamefont
  {Rudolph}},\ }\href {http://arxiv.org/abs/1607.08535
  https://arxiv.org/abs/1607.08535} {\bibfield  {journal} {\bibinfo  {journal}
  {arXiv:1607.08535}\ } (\bibinfo {year} {2016})},\ \Eprint
  {http://arxiv.org/abs/1607.08535} {arXiv:1607.08535} \BibitemShut {NoStop}%
\bibitem [{\citenamefont {Lindner}\ and\ \citenamefont
  {Rudolph}(2009)}]{2009.PRL.Lindner-Rudolph.ClusMachGun}%
  \BibitemOpen
  \bibfield  {author} {\bibinfo {author} {\bibfnamefont {N.~H.}\ \bibnamefont
  {Lindner}}\ and\ \bibinfo {author} {\bibfnamefont {T.}~\bibnamefont
  {Rudolph}},\ }\href {\doibase 10.1103/PhysRevLett.103.113602} {\bibfield
  {journal} {\bibinfo  {journal} {Physical Review Letters}\ }\textbf {\bibinfo
  {volume} {103}},\ \bibinfo {pages} {113602} (\bibinfo {year} {2009})},\
  \Eprint {http://arxiv.org/abs/0810.2587} {arXiv:0810.2587} \BibitemShut
  {NoStop}%
\bibitem [{\citenamefont {Schwartz}\ \emph {et~al.}(2016)\citenamefont
  {Schwartz}, \citenamefont {Cogan}, \citenamefont {Schmidgall}, \citenamefont
  {Don}, \citenamefont {Gantz}, \citenamefont {Kenneth}, \citenamefont
  {Lindner},\ and\ \citenamefont
  {Gershoni}}]{2016.Science.Schwartz-Gershoni.QDClusGen}%
  \BibitemOpen
  \bibfield  {author} {\bibinfo {author} {\bibfnamefont {I.}~\bibnamefont
  {Schwartz}}, \bibinfo {author} {\bibfnamefont {D.}~\bibnamefont {Cogan}},
  \bibinfo {author} {\bibfnamefont {E.~R.}\ \bibnamefont {Schmidgall}},
  \bibinfo {author} {\bibfnamefont {Y.}~\bibnamefont {Don}}, \bibinfo {author}
  {\bibfnamefont {L.}~\bibnamefont {Gantz}}, \bibinfo {author} {\bibfnamefont
  {O.}~\bibnamefont {Kenneth}}, \bibinfo {author} {\bibfnamefont {N.~H.}\
  \bibnamefont {Lindner}}, \ and\ \bibinfo {author} {\bibfnamefont
  {D.}~\bibnamefont {Gershoni}},\ }\href {\doibase 10.1126/science.aah4758}
  {\bibfield  {journal} {\bibinfo  {journal} {Science}\ }\textbf {\bibinfo
  {volume} {49}},\ \bibinfo {pages} {1804} (\bibinfo {year} {2016})},\ \Eprint
  {http://arxiv.org/abs/1606.07492} {arXiv:1606.07492} \BibitemShut {NoStop}%
\bibitem [{\citenamefont {Briegel}\ and\ \citenamefont
  {Raussendorf}(2001)}]{2001.PRL.Briegel-Raussendorf.ClustStateIntro}%
  \BibitemOpen
  \bibfield  {author} {\bibinfo {author} {\bibfnamefont {H.~J.}\ \bibnamefont
  {Briegel}}\ and\ \bibinfo {author} {\bibfnamefont {R.}~\bibnamefont
  {Raussendorf}},\ }\href {\doibase 10.1103/PhysRevLett.86.910} {\bibfield
  {journal} {\bibinfo  {journal} {Physical Review Letters}\ }\textbf {\bibinfo
  {volume} {86}},\ \bibinfo {pages} {910} (\bibinfo {year} {2001})}\BibitemShut
  {NoStop}%
\bibitem [{\citenamefont {Zhang}\ \emph {et~al.}(2008)\citenamefont {Zhang},
  \citenamefont {Bao}, \citenamefont {Lu}, \citenamefont {Zhou}, \citenamefont
  {Yang}, \citenamefont {Rudolph},\ and\ \citenamefont
  {Pan}}]{2008.PRA.Zhang-Pan.HeraldedBellPairsource}%
  \BibitemOpen
  \bibfield  {author} {\bibinfo {author} {\bibfnamefont {Q.}~\bibnamefont
  {Zhang}}, \bibinfo {author} {\bibfnamefont {X.-H.}\ \bibnamefont {Bao}},
  \bibinfo {author} {\bibfnamefont {C.-Y.}\ \bibnamefont {Lu}}, \bibinfo
  {author} {\bibfnamefont {X.-Q.}\ \bibnamefont {Zhou}}, \bibinfo {author}
  {\bibfnamefont {T.}~\bibnamefont {Yang}}, \bibinfo {author} {\bibfnamefont
  {T.}~\bibnamefont {Rudolph}}, \ and\ \bibinfo {author} {\bibfnamefont
  {J.-W.}\ \bibnamefont {Pan}},\ }\href {\doibase 10.1103/PhysRevA.77.062316}
  {\bibfield  {journal} {\bibinfo  {journal} {Physical Review A}\ }\textbf
  {\bibinfo {volume} {77}},\ \bibinfo {pages} {062316} (\bibinfo {year}
  {2008})}\BibitemShut {NoStop}%
\bibitem [{\citenamefont {Varnava}\ \emph {et~al.}(2008)\citenamefont
  {Varnava}, \citenamefont {Browne},\ and\ \citenamefont
  {Rudolph}}]{2008.PRL.Varnava-Rudolph.PhotonLoss-LOQCscaling}%
  \BibitemOpen
  \bibfield  {author} {\bibinfo {author} {\bibfnamefont {M.}~\bibnamefont
  {Varnava}}, \bibinfo {author} {\bibfnamefont {D.~E.}\ \bibnamefont {Browne}},
  \ and\ \bibinfo {author} {\bibfnamefont {T.}~\bibnamefont {Rudolph}},\ }\href
  {\doibase 10.1103/PhysRevLett.100.060502} {\bibfield  {journal} {\bibinfo
  {journal} {Physical Review Letters}\ }\textbf {\bibinfo {volume} {100}},\
  \bibinfo {pages} {060502} (\bibinfo {year} {2008})},\ \Eprint
  {http://arxiv.org/abs/0702044} {arXiv:0702044 [quant-ph]} \BibitemShut
  {NoStop}%
\bibitem [{\citenamefont {Michler}\ \emph {et~al.}(1996)\citenamefont
  {Michler}, \citenamefont {Mattle}, \citenamefont {Weinfurter},\ and\
  \citenamefont {Zeilinger}}]{1996.PRA.Michler-Zeilinger.BellStateAnalysis}%
  \BibitemOpen
  \bibfield  {author} {\bibinfo {author} {\bibfnamefont {M.}~\bibnamefont
  {Michler}}, \bibinfo {author} {\bibfnamefont {K.}~\bibnamefont {Mattle}},
  \bibinfo {author} {\bibfnamefont {H.}~\bibnamefont {Weinfurter}}, \ and\
  \bibinfo {author} {\bibfnamefont {A.}~\bibnamefont {Zeilinger}},\ }\href
  {\doibase 10.1103/PhysRevA.53.R1209} {\bibfield  {journal} {\bibinfo
  {journal} {Physical Review A}\ }\textbf {\bibinfo {volume} {53}},\ \bibinfo
  {pages} {R1209} (\bibinfo {year} {1996})}\BibitemShut {NoStop}%
\bibitem [{\citenamefont {Calsamiglia}\ and\ \citenamefont
  {L{\"{u}}tkenhaus}(2001)}]{2001.AppPhysB.Calsamiglia-Lutkenhaus.BSMlimit}%
  \BibitemOpen
  \bibfield  {author} {\bibinfo {author} {\bibfnamefont {J.}~\bibnamefont
  {Calsamiglia}}\ and\ \bibinfo {author} {\bibfnamefont {N.}~\bibnamefont
  {L{\"{u}}tkenhaus}},\ }\href {\doibase 10.1007/s003400000484} {\bibfield
  {journal} {\bibinfo  {journal} {Applied Physics B}\ }\textbf {\bibinfo
  {volume} {72}},\ \bibinfo {pages} {67} (\bibinfo {year} {2001})},\ \Eprint
  {http://arxiv.org/abs/0007058} {arXiv:0007058 [quant-ph]} \BibitemShut
  {NoStop}%
\bibitem [{\citenamefont
  {Hammersley}(1980)}]{1980.MathProcCambPhysSoc.Hammersley.sitebondperc}%
  \BibitemOpen
  \bibfield  {author} {\bibinfo {author} {\bibfnamefont {J.~M.}\ \bibnamefont
  {Hammersley}},\ }\href {\doibase 10.1017/S0305004100057455} {\bibfield
  {journal} {\bibinfo  {journal} {Mathematical Proceedings of the Cambridge
  Philosophical Society}\ }\textbf {\bibinfo {volume} {88}},\ \bibinfo {pages}
  {167} (\bibinfo {year} {1980})}\BibitemShut {NoStop}%
\bibitem [{\citenamefont {Tarasevich}\ and\ \citenamefont {van~der
  Marck}(1999)}]{1999.IJMPC.Tarasevich-vandermarck.sitebondapprox}%
  \BibitemOpen
  \bibfield  {author} {\bibinfo {author} {\bibfnamefont {Y.~Y.}\ \bibnamefont
  {Tarasevich}}\ and\ \bibinfo {author} {\bibfnamefont {S.~C.}\ \bibnamefont
  {van~der Marck}},\ }\href {\doibase 10.1142/S0129183199000978} {\bibfield
  {journal} {\bibinfo  {journal} {International Journal of Modern Physics C}\
  }\textbf {\bibinfo {volume} {10}},\ \bibinfo {pages} {14} (\bibinfo {year}
  {1999})},\ \Eprint {http://arxiv.org/abs/9906078} {arXiv:9906078 [cond-mat]}
  \BibitemShut {NoStop}%
\bibitem [{\citenamefont {Newman}\ and\ \citenamefont
  {Ziff}(2001)}]{2001.PRE.Newman-Ziff.NZAlgo}%
  \BibitemOpen
  \bibfield  {author} {\bibinfo {author} {\bibfnamefont {M.~E.~J.}\
  \bibnamefont {Newman}}\ and\ \bibinfo {author} {\bibfnamefont {R.~M.}\
  \bibnamefont {Ziff}},\ }\href {\doibase 10.1103/PhysRevE.64.016706}
  {\bibfield  {journal} {\bibinfo  {journal} {Physical Review E - Statistical,
  Nonlinear, and Soft Matter Physics}\ }\textbf {\bibinfo {volume} {64}},\
  \bibinfo {pages} {1} (\bibinfo {year} {2001})},\ \Eprint
  {http://arxiv.org/abs/0101295} {arXiv:0101295 [cond-mat]} \BibitemShut
  {NoStop}%
\bibitem [{\citenamefont {Tran}\ \emph {et~al.}(2013)\citenamefont {Tran},
  \citenamefont {Yoo}, \citenamefont {Stahlheber},\ and\ \citenamefont
  {Small}}]{2013.JSMTE.Tran-Small.3Dlattdeg3perc}%
  \BibitemOpen
  \bibfield  {author} {\bibinfo {author} {\bibfnamefont {J.}~\bibnamefont
  {Tran}}, \bibinfo {author} {\bibfnamefont {T.}~\bibnamefont {Yoo}}, \bibinfo
  {author} {\bibfnamefont {S.}~\bibnamefont {Stahlheber}}, \ and\ \bibinfo
  {author} {\bibfnamefont {A.}~\bibnamefont {Small}},\ }\href {\doibase
  10.1088/1742-5468/2013/05/P05014} {\bibfield  {journal} {\bibinfo  {journal}
  {Journal of Statistical Mechanics: Theory and Experiment}\ }\textbf {\bibinfo
  {volume} {2013}},\ \bibinfo {pages} {P05014} (\bibinfo {year}
  {2013})}\BibitemShut {NoStop}%
\bibitem [{\citenamefont {{Van Den Nest}}\ \emph {et~al.}(2006)\citenamefont
  {{Van Den Nest}}, \citenamefont {Miyake}, \citenamefont {D{\"{u}}r},\ and\
  \citenamefont {Briegel}}]{2006.PRL.VanDenNest-Briegel.UnivResMBQC}%
  \BibitemOpen
  \bibfield  {author} {\bibinfo {author} {\bibfnamefont {M.}~\bibnamefont {{Van
  Den Nest}}}, \bibinfo {author} {\bibfnamefont {A.}~\bibnamefont {Miyake}},
  \bibinfo {author} {\bibfnamefont {W.}~\bibnamefont {D{\"{u}}r}}, \ and\
  \bibinfo {author} {\bibfnamefont {H.~J.}\ \bibnamefont {Briegel}},\ }\href
  {\doibase 10.1103/PhysRevLett.97.150504} {\bibfield  {journal} {\bibinfo
  {journal} {Physical Review Letters}\ }\textbf {\bibinfo {volume} {97}}
  (\bibinfo {year} {2006}),\ 10.1103/PhysRevLett.97.150504},\ \Eprint
  {http://arxiv.org/abs/0604010} {arXiv:0604010 [quant-ph]} \BibitemShut
  {NoStop}%
\bibitem [{\citenamefont {Varnava}\ \emph {et~al.}(2006)\citenamefont
  {Varnava}, \citenamefont {Browne},\ and\ \citenamefont
  {Rudolph}}]{2006.PRL.Varnava-Rudolph.CountEC}%
  \BibitemOpen
  \bibfield  {author} {\bibinfo {author} {\bibfnamefont {M.}~\bibnamefont
  {Varnava}}, \bibinfo {author} {\bibfnamefont {D.}~\bibnamefont {Browne}}, \
  and\ \bibinfo {author} {\bibfnamefont {T.}~\bibnamefont {Rudolph}},\ }\href
  {\doibase 10.1103/PhysRevLett.97.120501} {\bibfield  {journal} {\bibinfo
  {journal} {Physical Review Letters}\ }\textbf {\bibinfo {volume} {97}},\
  \bibinfo {pages} {120501} (\bibinfo {year} {2006})}\BibitemShut {NoStop}%
\bibitem [{\citenamefont {Azuma}\ \emph {et~al.}(2015)\citenamefont {Azuma},
  \citenamefont {Tamaki},\ and\ \citenamefont
  {Lo}}]{2015.NatureComm.Azuma-Lo.AllOptRep}%
  \BibitemOpen
  \bibfield  {author} {\bibinfo {author} {\bibfnamefont {K.}~\bibnamefont
  {Azuma}}, \bibinfo {author} {\bibfnamefont {K.}~\bibnamefont {Tamaki}}, \
  and\ \bibinfo {author} {\bibfnamefont {H.-K.}\ \bibnamefont {Lo}},\ }\href
  {\doibase 10.1038/ncomms7787} {\bibfield  {journal} {\bibinfo  {journal}
  {Nature communications}\ }\textbf {\bibinfo {volume} {6}},\ \bibinfo {pages}
  {6787} (\bibinfo {year} {2015})}\BibitemShut {NoStop}%
\bibitem [{\citenamefont {Pant}\ \emph {et~al.}(2017)\citenamefont {Pant},
  \citenamefont {Krovi}, \citenamefont {Englund},\ and\ \citenamefont
  {Guha}}]{2017.PRA.Pant-Guha.AllOptRepRescources}%
  \BibitemOpen
  \bibfield  {author} {\bibinfo {author} {\bibfnamefont {M.}~\bibnamefont
  {Pant}}, \bibinfo {author} {\bibfnamefont {H.}~\bibnamefont {Krovi}},
  \bibinfo {author} {\bibfnamefont {D.}~\bibnamefont {Englund}}, \ and\
  \bibinfo {author} {\bibfnamefont {S.}~\bibnamefont {Guha}},\ }\href {\doibase
  10.1103/PhysRevA.95.012304} {\bibfield  {journal} {\bibinfo  {journal}
  {Physical Review A}\ }\textbf {\bibinfo {volume} {95}},\ \bibinfo {pages}
  {012304} (\bibinfo {year} {2017})},\ \Eprint
  {http://arxiv.org/abs/1603.01353} {arXiv:1603.01353} \BibitemShut {NoStop}%
\bibitem [{\citenamefont {Broadbent}\ and\ \citenamefont
  {Hammersley}(1957)}]{1957.MathProcCambPhilSoc.Broadbent-Hammersley.Percolation}%
  \BibitemOpen
  \bibfield  {author} {\bibinfo {author} {\bibfnamefont {S.~R.}\ \bibnamefont
  {Broadbent}}\ and\ \bibinfo {author} {\bibfnamefont {J.~M.}\ \bibnamefont
  {Hammersley}},\ }\href {\doibase 10.1017/} {\bibfield  {journal} {\bibinfo
  {journal} {Mathematical Proceedings of the Cambridge Philosophical Society}\
  }\textbf {\bibinfo {volume} {53}},\ \bibinfo {pages} {629} (\bibinfo {year}
  {1957})}\BibitemShut {NoStop}%
\bibitem [{\citenamefont {Sahimi}\ \emph {et~al.}(1983)\citenamefont {Sahimi},
  \citenamefont {Hughes}, \citenamefont {Scriven},\ and\ \citenamefont
  {Davis}}]{1983.JPAMG.Sahimi-Davis.BPthresh}%
  \BibitemOpen
  \bibfield  {author} {\bibinfo {author} {\bibfnamefont {M.}~\bibnamefont
  {Sahimi}}, \bibinfo {author} {\bibfnamefont {B.~D.}\ \bibnamefont {Hughes}},
  \bibinfo {author} {\bibfnamefont {L.~E.}\ \bibnamefont {Scriven}}, \ and\
  \bibinfo {author} {\bibfnamefont {H.~T.}\ \bibnamefont {Davis}},\ }\href
  {\doibase 10.1088/0305-4470/16/2/004} {\bibfield  {journal} {\bibinfo
  {journal} {J. Phys. A: Math. Gen.}\ }\textbf {\bibinfo {volume} {16}},\
  \bibinfo {pages} {67} (\bibinfo {year} {1983})}\BibitemShut {NoStop}%
\bibitem [{\citenamefont {Gimeno-Segovia}(2015)}]{2015.PhDThesis.Mercedes}%
  \BibitemOpen
  \bibfield  {author} {\bibinfo {author} {\bibfnamefont {M.}~\bibnamefont
  {Gimeno-Segovia}},\ }\emph {\bibinfo {title} {{Towards practical Linear
  Optical Quantum Computing}}},\ \href@noop {} {Ph.D. thesis},\ \bibinfo
  {school} {Imperial College London} (\bibinfo {year} {2015})\BibitemShut
  {NoStop}%
\bibitem [{\citenamefont {Varnava}\ \emph {et~al.}(2007)\citenamefont
  {Varnava}, \citenamefont {Browne},\ and\ \citenamefont
  {Rudolph}}]{2007.NJP.Varnava-Rudolph.CountFactECMemory}%
  \BibitemOpen
  \bibfield  {author} {\bibinfo {author} {\bibfnamefont {M.}~\bibnamefont
  {Varnava}}, \bibinfo {author} {\bibfnamefont {D.~E.}\ \bibnamefont {Browne}},
  \ and\ \bibinfo {author} {\bibfnamefont {T.}~\bibnamefont {Rudolph}},\ }\href
  {\doibase 10.1088/1367-2630/9/6/203} {\bibfield  {journal} {\bibinfo
  {journal} {New Journal of Physics}\ }\textbf {\bibinfo {volume} {9}},\
  \bibinfo {pages} {203} (\bibinfo {year} {2007})}\BibitemShut {NoStop}%
\bibitem [{\citenamefont {Barrett}\ and\ \citenamefont
  {Stace}(2010)}]{2010.PRL.Barrett-Stace.TopolPhotonLoss}%
  \BibitemOpen
  \bibfield  {author} {\bibinfo {author} {\bibfnamefont {S.~D.}\ \bibnamefont
  {Barrett}}\ and\ \bibinfo {author} {\bibfnamefont {T.~M.}\ \bibnamefont
  {Stace}},\ }\href {\doibase 10.1103/PhysRevLett.105.200502} {\bibfield
  {journal} {\bibinfo  {journal} {Physical Review Letters}\ }\textbf {\bibinfo
  {volume} {105}},\ \bibinfo {pages} {200502} (\bibinfo {year}
  {2010})}\BibitemShut {NoStop}%
\bibitem [{\citenamefont {Carolan}\ \emph {et~al.}(2015)\citenamefont
  {Carolan}, \citenamefont {Harrold}, \citenamefont {Sparrow}, \citenamefont
  {Martin-Lopez}, \citenamefont {Russell}, \citenamefont {Silverstone},
  \citenamefont {Shadbolt}, \citenamefont {Matsuda}, \citenamefont {Oguma},
  \citenamefont {Itoh}, \citenamefont {Marshall}, \citenamefont {Thompson},
  \citenamefont {Matthews}, \citenamefont {Hashimoto}, \citenamefont
  {O'Brien},\ and\ \citenamefont {Laing}}]{2015.Science.Carolan-Laing.UnivLO}%
  \BibitemOpen
  \bibfield  {author} {\bibinfo {author} {\bibfnamefont {J.}~\bibnamefont
  {Carolan}}, \bibinfo {author} {\bibfnamefont {C.}~\bibnamefont {Harrold}},
  \bibinfo {author} {\bibfnamefont {C.}~\bibnamefont {Sparrow}}, \bibinfo
  {author} {\bibfnamefont {E.}~\bibnamefont {Martin-Lopez}}, \bibinfo {author}
  {\bibfnamefont {N.~J.}\ \bibnamefont {Russell}}, \bibinfo {author}
  {\bibfnamefont {J.~W.}\ \bibnamefont {Silverstone}}, \bibinfo {author}
  {\bibfnamefont {P.~J.}\ \bibnamefont {Shadbolt}}, \bibinfo {author}
  {\bibfnamefont {N.}~\bibnamefont {Matsuda}}, \bibinfo {author} {\bibfnamefont
  {M.}~\bibnamefont {Oguma}}, \bibinfo {author} {\bibfnamefont
  {M.}~\bibnamefont {Itoh}}, \bibinfo {author} {\bibfnamefont {G.~D.}\
  \bibnamefont {Marshall}}, \bibinfo {author} {\bibfnamefont {M.~G.}\
  \bibnamefont {Thompson}}, \bibinfo {author} {\bibfnamefont {J.~C.~F.}\
  \bibnamefont {Matthews}}, \bibinfo {author} {\bibfnamefont {T.}~\bibnamefont
  {Hashimoto}}, \bibinfo {author} {\bibfnamefont {J.~L.}\ \bibnamefont
  {O'Brien}}, \ and\ \bibinfo {author} {\bibfnamefont {A.}~\bibnamefont
  {Laing}},\ }\href {\doibase 10.1126/science.aab3642} {\bibfield  {journal}
  {\bibinfo  {journal} {Science}\ }\textbf {\bibinfo {volume} {349}},\ \bibinfo
  {pages} {711} (\bibinfo {year} {2015})},\ \Eprint
  {http://arxiv.org/abs/1505.01182} {arXiv:1505.01182} \BibitemShut {NoStop}%
\bibitem [{\citenamefont {Mower}\ \emph {et~al.}(2015)\citenamefont {Mower},
  \citenamefont {Harris}, \citenamefont {Steinbrecher}, \citenamefont
  {Lahini},\ and\ \citenamefont {Englund}}]{2015.PRA.Mower-Englund.QPPTheory}%
  \BibitemOpen
  \bibfield  {author} {\bibinfo {author} {\bibfnamefont {J.}~\bibnamefont
  {Mower}}, \bibinfo {author} {\bibfnamefont {N.~C.}\ \bibnamefont {Harris}},
  \bibinfo {author} {\bibfnamefont {G.~R.}\ \bibnamefont {Steinbrecher}},
  \bibinfo {author} {\bibfnamefont {Y.}~\bibnamefont {Lahini}}, \ and\ \bibinfo
  {author} {\bibfnamefont {D.}~\bibnamefont {Englund}},\ }\href {\doibase
  10.1103/PhysRevA.92.032322} {\bibfield  {journal} {\bibinfo  {journal}
  {Physical Review A - Atomic, Molecular, and Optical Physics}\ }\textbf
  {\bibinfo {volume} {92}} (\bibinfo {year} {2015}),\
  10.1103/PhysRevA.92.032322},\ \Eprint {http://arxiv.org/abs/1406.3255}
  {arXiv:1406.3255} \BibitemShut {NoStop}%
\bibitem [{\citenamefont {Harris}\ \emph {et~al.}(2015)\citenamefont {Harris},
  \citenamefont {Steinbrecher}, \citenamefont {Mower}, \citenamefont {Lahini},
  \citenamefont {Prabhu}, \citenamefont {Baehr-Jones}, \citenamefont
  {Hochberg}, \citenamefont {Lloyd},\ and\ \citenamefont
  {Englund}}]{2015.arXiv.Harris-Englund.BosTransQPP}%
  \BibitemOpen
  \bibfield  {author} {\bibinfo {author} {\bibfnamefont {N.~C.}\ \bibnamefont
  {Harris}}, \bibinfo {author} {\bibfnamefont {G.~R.}\ \bibnamefont
  {Steinbrecher}}, \bibinfo {author} {\bibfnamefont {J.}~\bibnamefont {Mower}},
  \bibinfo {author} {\bibfnamefont {Y.}~\bibnamefont {Lahini}}, \bibinfo
  {author} {\bibfnamefont {M.}~\bibnamefont {Prabhu}}, \bibinfo {author}
  {\bibfnamefont {T.}~\bibnamefont {Baehr-Jones}}, \bibinfo {author}
  {\bibfnamefont {M.}~\bibnamefont {Hochberg}}, \bibinfo {author}
  {\bibfnamefont {S.}~\bibnamefont {Lloyd}}, \ and\ \bibinfo {author}
  {\bibfnamefont {D.}~\bibnamefont {Englund}},\ }\href
  {http://arxiv.org/abs/1507.03406} {\bibfield  {journal} {\bibinfo  {journal}
  {arXiv: 1507.03406v2}\ }\textbf {\bibinfo {volume} {24}},\ \bibinfo {pages}
  {1} (\bibinfo {year} {2015})},\ \Eprint {http://arxiv.org/abs/1507.03406}
  {arXiv:1507.03406} \BibitemShut {NoStop}%
\end{thebibliography}%

\end{document}